\documentclass[conference,compsoc]{IEEEtran}
\IEEEoverridecommandlockouts

\usepackage[
    style=ieee,
    natbib=true,
    backend=biber,
]{biblatex}

\usepackage[utf8]{inputenc} 
\usepackage[T1]{fontenc} 

\usepackage{graphics}
\usepackage[dvipsnames]{xcolor}

\usepackage{amsmath,amssymb,amsfonts}
\usepackage{nicefrac}       

\usepackage{booktabs} 
\usepackage{multirow}
\usepackage{multicol}

\usepackage[hidelinks]{hyperref}
\usepackage{url}            

\usepackage{csquotes}
\usepackage{pifont}
\usepackage{textcomp}

\usepackage{caption, subcaption}

\usepackage[normalem]{ulem}

\usepackage[acronym,nohypertypes={acronym,notation}]{glossaries}

\usepackage{enumitem}

\usepackage{microtype}      

\usepackage[nolist,nohyperlinks]{acronym}

 \usepackage{todonotes}

\usepackage[framemethod=TikZ]{mdframed}
\definecolor{rubcolor}{HTML}{093158}
\global\mdfdefinestyle{insightstyle}{backgroundcolor=rubcolor!2,
outerlinewidth=1pt,innerlinewidth=0pt,
outerlinecolor=rubcolor,roundcorner=5pt
}
\newmdenv[roundcorner=3pt, nobreak=true, innertopmargin=7pt, innerbottommargin=8pt,skipabove=12pt, linecolor=rubcolor, backgroundcolor=rubcolor!10]{insightbox}

\usepackage{wasysym} 

\definecolor{myblue}{HTML}{10589B}

\newcommand{\cf}{cf.}

\newcommand{\eg}{e.g.}
\newcommand{\ie}{i.e.}

\newtheorem{finding}{Finding}

\def\BibTeX{{\rm B\kern-.05em{\sc i\kern-.025em b}\kern-.08em
    T\kern-.1667em\lower.7ex\hbox{E}\kern-.125emX}}
\begin{document}

\begin{acronym}
    \acro{GTS}{Generalized Trust Score}
    \acro{FAM}{Familiarity}
    \acro{PI}{Prediction Interval}
    \acro{CRT}{Cognitive Reflection Test}
    \acro{AHS}{Analysis-Holism Scale}
    \acro{NMLS}{New Media Literacy Scale}
    \acro{PO}{Political Orientation}
    \acro{TTS}{\emph{Text-To-Speech}}
\end{acronym}

\newcommand{\gts}{\ac{GTS}}
\newcommand{\pred}{\ac{PI}}
\newcommand{\fami}{\ac{FAM}}
\newcommand{\crt}{\ac{CRT}}
\newcommand{\ahs}{\ac{AHS}}
\newcommand{\nmls}{\ac{NMLS}}
\newcommand{\po}{\ac{PO}}
\newcommand{\tts}{\ac{TTS}}

\title{{A Representative Study on Human Detection \\ of Artificially Generated Media Across Countries}
}

\author{
\IEEEauthorblockN{Joel Frank}
\IEEEauthorblockA{
\textit{Ruhr-Universität Bochum} \\
joel.frank@rub.de}
\and
\IEEEauthorblockN{Franziska Herbert}
\IEEEauthorblockA{
\textit{Ruhr-Universität Bochum} \\
franziska.herbert@rub.de}
\and
\IEEEauthorblockN{Jonas Ricker}
\IEEEauthorblockA{
\textit{Ruhr-Universität Bochum} \\
\phantom{c.uni-han}jonas.ricker@rub.de\phantom{nover.de}}
\and
\IEEEauthorblockN{Lea Schönherr}
\IEEEauthorblockA{
\textit{CISPA} \\
schoenherr@cispa.de}
\and
\IEEEauthorblockN{Thorsten Eisenhofer}
\IEEEauthorblockA{
\textit{TU Berlin} \\
thorsten.eisenhofer@tu-berlin.de} \and
\IEEEauthorblockN{Asja Fischer}
\IEEEauthorblockA{
\textit{Ruhr-Universität Bochum} \\
asja.fischer@rub.de}
\and
\IEEEauthorblockN{Markus Dürmuth}
\IEEEauthorblockA{
\textit{Leibniz Universität Hannover} \\
markus.duermuth@itsec.uni-hannover.de}
\and
\IEEEauthorblockN{Thorsten Holz}
\IEEEauthorblockA{
\textit{CISPA} \\
holz@cispa.de}
}

\maketitle

\begin{abstract}
AI-generated media has become a threat to our digital society as we know it.
Forgeries can be created automatically and on a large scale based on publicly available technologies.
Recognizing this challenge, academics and practitioners have proposed a multitude of automatic detection strategies to detect such artificial media.
However, in contrast to these technological advances, the \emph{human perception} of generated media has not been thoroughly studied yet.

In this paper, we aim to close this research gap.
We conduct the first comprehensive survey on people's ability to detect generated media, spanning three countries (USA, Germany, and China), with 3,002 participants covering audio, image, and text media.
Our results indicate that state-of-the-art forgeries are almost indistinguishable from ``real'' media, with the majority of participants simply guessing when asked to rate them as human- or machine-generated.
In addition, AI-generated media is rated as more likely to be human-generated across all media types and all countries.
To further understand which factors influence people's ability to detect AI-generated media, we include personal variables, chosen based on a literature review in the domains of deepfake and fake news research.
In a regression analysis, we found that generalized trust, cognitive reflection, and self-reported familiarity with deepfakes significantly influence participants' decisions across all media categories.
\end{abstract}

\section{Introduction}
\label{sec:intro}

In his 2015 book, the historian Yuval Noah Harari wrote: ``In the past, censorship worked by blocking the flow of information. In the 21st century, censorship works by flooding people with irrelevant information.''~\citep{harari2016homo}.
While in the original context the quote referred to \emph{fake news}, the content of the quote is more relevant than ever.
Deep generative modeling has enabled the unbounded creation of fake media at scale.
While it has been used for harmless endeavors like putting Jim Carrey into  the movie \emph{Shining}~\citep{sharf2019jimcarry}, there are also more destructive examples like phishing \$243,000 from a UK company by imitating their CEO's voice~\citep{stupp2019ceo} or influencing political events~\citep{thompson2017memewarfare,hao2019deepfake,mwai2021tigray,dfrlab2021navalny}. A prominent example being a fake video of Ukraine's president Zelenskyy telling his forces to surrender~\citep{wakefield2022ukraine}.
These fakes have resulted in a flurry of techniques to automatically detect AI-generated media~\citep{zhang2019detecting,qian2020thinking,yu2019attributing,frank2020leveraging,marra2019gans,wang2019cnn,tariq2019gan,mccloskey2018detecting,nataraj2019detecting,mo2018fake,marra2018detection,frank2021wavefake,ricker2022towards, pu-22-text-detection}.
However, there is little research on how convincing AI-generated media is to human observers. 
Prior works~\citep{hulzebosch2020detecting,hwang2021effects,minkdeepphish,nightingaleAIsynthesizedFacesAre2022,lago2022more,muller2022human} often only consider one type of media (usually images) and often rely on small sample sizes or convenience samples.

In this work, we establish the first cross-country and cross-media baseline the detection of media generated with state-of-the-art methods.
In our preregistered study, we examine three different countries (USA, Germany, and China) under three different conditions (audio, image, and text), with a total number of participation of $n=3,002$.
The primary goal of the surveys is to answer the following three research questions: i) Can people identify state-of-the-art generated media? ii) Which demographic factors influence the identification accuracy? iii) Which cognitive factors affect the identification accuracy?

In our survey, participants are asked to rate a set of human- and machine-generated media on how believable they are.
Most importantly, we find that most AI-generated samples are already so convincing that the majority of participants cannot accurately distinguish them from human-generated content.
More specifically, the average detection accuracy of participants is below 50\% for images and never exceeds 60\% for the other media types.
Moreover, we find that participants in all countries believed that most of the samples we showed to them were human-generated, compared to the $50/50$ ground truth.

To evaluate which variables might improve or worsen people's ability to detect AI-generated media, we performed a literature review of prior work in the domains of deepfake and fake news research.
This review enabled us to identify multiple personal variables that might influence the decision of participants, such as cognitive reasoning~\citep{Appel2022}, media literacy~\citep{hwang2021effects} or political orientation~\citep{AHMED2021101508}. We included these variables in our survey.
In a regression analysis, we found that generalized trust, motivated System 2 reasoning, and self-reported familiarity with deepfakes significantly influenced people's decision across all media categories.
Furthermore, we found several other variables that influence the decision dependent on the media type.

\smallskip \noindent
In summary, we make the following key contributions:

\begin{itemize}
    \item We conduct the first preregistered cross-country and cross-media survey regarding AI-generated media with more than 3,000 participants.
    \item Our results indicate that AI-generated media is already so convincing that the majority of participants simply guess when asked to rate them as human- or machine-generated media.
    \item Using a regression analysis, we show a significant influence of generalized trust, motivated System 2 reasoning, and self-reported familiarity with deepfakes across all our conditions.
    Additionally, we found several condition-dependent influential factors.
\end{itemize}

All information regarding the preregistration is available at \href{https://osf.io/xy6v5}{https://osf.io/xy6v5}. 
The code to conduct the study and all of our analysis can be found online at \href{https://github.com/RUB-SysSec/GeneratedMediaSurvey}{github.com/RUB-SysSec/GeneratedMediaSurvey}. \section{Related Work}
\label{sec:related}

Due to the lack of comprehensive cross-media and cross-country prior work, we survey related work in the field of AI-generated media and fake news.
This allows us to connect our work to prior work focusing on specific kinds of AI-generated media (e.g., images) and to comment on the transferability of methods and techniques developed in adjacent domains.
Additionally, we provide an overview of the current state-of-the-art in generative modeling.

\subsection{Generative Modeling}
\label{sec:related:generative}
Generative modeling---the technology behind artificially generated media---has received tremendous attention in recent years:
First, Generative Adversarial Networks (GANs)~\citep{goodfellow2014generative} started a wave of publications~\citep{radford2015unsupervised,mirza2014conditional,salimans2016improved,arjovsky2017wasserstein,gulrajani2017improved,petzka2017regularization,miyato2018spectral,karras2018progressive,karras2019style,brock2018large,Karras_2020_CVPR}, the most prominent being several iterations of a model called StyleGAN, which, for the first time, generated photo-realistic portraits of human faces.
While GANs had a profound impact, recently, the focus of the image domain shifted to diffusion models~\citep{ho2020denoising,nichol2021improved,dhariwal2021diffusion}, the most famous being StableDiffusion~\citep{rombach2021highresolution} and Dall-E~2~\citep{ramesh2022hierarchical}.
These models can generate thousands of different image variants for simple text prompts with the help of large language models.
Large language models are a different kind of generative models focused on generating legible text that appears to be created by humans.
These models have their origin in machine translation~\citep{bahdanau2014neural,cho2014learning,sutskever2014sequence}, a subtask of generative modeling, where models were prompted with a paragraph in one language (\eg, English) and had to produce the equivalent in another language (\eg, French).
The big breakthrough was the introduction of the attention concept~\citep{bahdanau2014neural,kim2017structured} and the corresponding architecture called Transformers~\citep{vaswani2017attention}.
Today, Transformer-based architectures like GPT-3~\citep{brown-20-language}/GPT-4~\cite{openai-gpt4} or PaLM~\citep{chowdhery2022palm} are prompted with small summaries of text and generate entire paragraphs expanding on the prompt.
Finally, the synthetic speech landscape has been the most recent to be completely transformed by advances in deep learning.
While traditional approaches used hand-crafted algorithms imitating human speech patterns~\citep{zen2009statistical,tokuda2000speech,yoshimura1999simultaneous}, today's algorithms use a combination of two neural networks to generate human speech from text prompts~\citep{kumar2019melgan,binkowski2019high,donahue2018adversarial,kong2020hifi,yamamoto2020pwg}, so-called \tts{} models.

\subsection{Personal Variables}
\label{sec:related:variables}
We have organized this section according to different personal variables that have been found to influence people's decisions.

\smallskip

\textbf{Media Literacy}. 
\label{sec:related:nmls}
Previous work has suggested a link between media literacy and the susceptibility to fake information or disinformation~\citep{rubin2019disinformation,jang2018third,jones2021does}.
The common assumption is that those individuals with a higher media literacy engage more critical in the media they consume.
For example, a recent meta-analytic study by~\citet{jeong2012media} indicates that people with a better understanding of media and media production systems tend to be more skeptical and realistic about media messages.
Prior work has already linked the media literacy interventions to a decreased willingness to share deepfake videos~\citep{hwang2021effects}.

\textbf{Holistic Thinking}.
\label{sec:related:ahs}
People from East Asian cultures tend to think more holistically, while people from Western cultures tend to think more analytically~\citep{munro1985individualism,nakamura1991ways,masuda2001attending,nisbett2001culture}.
East Asians focus more on the relationship between objects and the field to which it belongs.
In contrast, Westerners apply a more analytic style, focusing their attention more on an object itself~\citep{nisbett2001culture}.
In the realm of fake news, previous work has shown a negative correlation between analytic thinking and perceived accuracy of fake news~\citep{pennycook2020falls,bronstein2019belief}.

\textbf{Generalized Trust}.
\label{sec:related:trust}
Trust is a core element of society, and everyday social life would be impossible without it~\citep{helliwell2006well,putnam2000bowling}.
It has been defined as a willingness to be vulnerable to the actions of others~\citep{mayer1995integrative} and arises from social attitudes regarding the world and other people.
These attitudes can be developed either in a general or interpersonal context~\citep{putnam2000bowling}.
General or depersonalized trust refers to one's trust towards public institutions or out-group members~\citep{maddux2005gender}.
Past research has found connections between trust and favorable national outcomes, such as economic growth and earnings, as well as a variety of desirable interpersonal qualities (\eg, social solidarity, tolerance, volunteerism, cooperation, optimism;~\citep{ashraf2006decomposing,rothstein2005all,tov2009well}).
It is also crucial for online social interactions like online dating~\citep{gibbs2011first,ma2017happens} or marketplaces like Airbnb~\citep{ert2016trust,lampinen2016hosting}.
In the context of machine-generated media,~\citet{minkdeepphish} have shown that people with a higher general trust are more likely to trust and accept friend requests from machine-generated LinkedIn profiles.

\textbf{Cognitive Reflection}.
\label{sec:related:CRT}
The dual process theory is a cognitive framework that proposes two distinct modes of thinking: \emph{System 1} and \emph{System 2}.
System 1 acts automatically and spontaneously and does not require conscious reflection, while System 2 is believed to require deliberation, analytic thinking, and concentration~\citep{stanovich_west_2000,frederick2005cognitive}.
We hypothesize that people who engage more in System 2 reasoning might be better at recognizing deepfakes because they take a more conscious approach to evaluating media instead of relying on gut decisions.
We administer the \crt{}~\citep{frederick2005cognitive}, which measures a person's ability to engage in System 2 reasoning.
Previous work found evidence that achieving a high score on the \crt{} is positively correlated to correctly discerning fake from real news~\citep{bronstein2019belief,PENNYCOOK201939,pennycook2020falls,Bago2020,PENNYCOOK2021388,Batailler2022}.
Moreover, prior work showed that people are not more likely to believe fake news that are consistent with their own political ideology~\citep{PENNYCOOK201939,Bago2020,Batailler2022}.
In other words, people fall for fake news because of ``lazy thinking'' and not because it corresponds to their beliefs~\citep{PENNYCOOK201939}.
A real-world analysis of Twitter data also suggests that people with higher \crt{} scores are more likely to share high-quality content and therefore do not support the dissemination of fake news~\citep{Mosleh2021}.
For deepfakes, previous works have made similar findings: Cognitive reflection is positively correlated to the ability to detect artificially generated media~\citep{Groh2022,Appel2022} and negatively correlated to inadvertent sharing of deepfakes~\citep{AHMED2021101508,AHMED2021111074}.
However, in a study including a political deepfake video, \citet{Appel2022} have shown that although a higher \crt{} score helps identify implausible content, it does not have an influence on the ability to spot generation artifacts (\eg, glitches or out-of-sync lips).

\textbf{Political Orientation}.
\label{sec:related:PO}
Previous work suggests a link between political interest and orientation, and fake news and deepfake engagement (see~\citep{Calvillo_2020,AHMED2021101508,chadwick2019news}). For example, participants with higher political interest from both the US and Singapore were more likely to unintentionally share deepfakes~\citep{AHMED2021101508}. More exposure to deepfakes was associated with being more likely to share deepfakes inadvertently, with US participants reporting more exposure to deepfakes than participants from Singapore. \citet{chadwick2019news} also found that people in the UK with a higher political interest were more likely to share exaggerated or false information on social media, both intentionally and unintentionally. In one of two studies with deepfake video(s) of politicians by~\citet{Appel2022}, political interest was positively associated with the likelihood of detecting the deepfake. However, in the second study this effect was not found~\citep{Appel2022}. In addition to (increased) political interest, conservatism was found to be related to fake news identification. \citet{Calvillo_2020} found that conservatism and news discernment, in this case headlines related to COVID-19, correlated negatively.

 \section{Method}
\label{sec:method}
To obtain a comprehensive overview of people's ability to detect deepfakes, we conducted an online survey in the USA, Germany, and China between June and September 2022.
These three countries were chosen for three reasons: i) availability of high-quality generative models, ii) the ability to obtain a sufficiently large qualitative sample, and iii) a diversity of cultures.
To conduct the survey, we assigned the participants to three groups and confronted them with either audio files, images, or texts.
We then asked our participants to rate each stimulus as human- or machine-generated.
We have preregistered the whole study design, including the sampling and analysis plan, via the Open Science Framework (OSF) before conducting the study.

\subsection{AI-generated Media}
\label{sec:method:generation}

As the basis for our study, we collect and generate human- and machine-generated media for the three considered conditions. In the following, we describe the conducted process to obtain the audio, image, and text stimuli. Examples of all media types can be found in Appendix~\ref{app:stimuli}.

\subsubsection{Audio} 
Speech samples are generated by a text-to-speech (TTS) pipeline based on two components: the acoustic model, specifically a Tacotron~2 model~\cite{shen2018natural}, and the vocoder, in form of Hifi-GAN~\cite{kong2020hifi}.
Both are state-of-the-art models used in current TTS pipelines.
The first component transforms text into a Mel spectrogram representation, and the second component creates a raw waveform using the Mel spectrogram as input. For each language, specifically trained models are utilized, using the following three datasets: For English we use the LJSpeech dataset~\cite{ljspeech17}, for Chinese the CSMSC dataset~\cite{csmsc17}, and for German the HUI dataset~\cite{puchtler2021hui}.
For the generated files, we randomly picked a disjoint set of 15 samples and generated corresponding machine-generated versions with the generative models for the respective language, using the original text as input.
For the human comparison, we randomly picked another 15 samples from the corresponding training set.

\subsubsection{Image}
\label{sec:method:generation:images}
We use a subset of the real and machine-generated images used by Nightingale and Farid~\citep{nightingaleAIsynthesizedFacesAre2022}.
They generated synthetic faces using StyleGAN2~\citep{Karras_2020_CVPR} and created a collection of 400 images that are equally split across gender, age, and ethnicity (African American or Black, Caucasian, East Asian, South Asian).
Additionally, they manually checked synthetic images for uniform backgrounds and no obvious rendering artifacts.
Each artificial image was matched to a real image from the training dataset of StyleGAN2, using its feature representation obtained by a deep convolutional neural network~\citep{parkhi2015a}.
Specifically, the network extracted a 4,096-dimensional vector, which was then compared to those of all 70,000 real faces using Euclidean distance.
They manually picked the most similar image from the best matches, taking position, posture, facial expression, and presence of accessories into account.
For this survey, we picked 16 pairs of real and machine-generated images, ensuring an equal distribution of gender and ethnicity.

\subsubsection{Text} 
We generate fake news articles with state-of-the-art language models for text generation.
Real articles are collected from a major press agency in the respective country: National Public Radio (NPR) \cite{misc-22-npr} in the USA, Tagesschau \cite{misc-22-tagesschau} in Germany, and China Central Television (CCTV) \cite{misc-22-cctv} in China. All agencies are rated with a neutral media bias \cite{misc-22-allsides} (\ie, they are neither left or right oriented). To make news articles comparable across countries, we consider articles from the business, national, and international category.

For generation, we follow prior work and use a few-shot learning approach \cite{brown-20-language}, in which a pre-trained model is fine-tuned for a specific task by presenting it with few examples of the desired output (\ie, a news article).
We generate a sample by prompting the model with two news articles composed of a title, short summary, and the main text.
The final sample is then generated by providing the model only with the title and summary of an article and outputting the main text by the model.
We use the Davinci GPT3 model from OpenAI \cite{misc-22-openai}, which is capable of generating English, German, as well as Chinese text. We configure the model with its default parameters. To obtain a larger variety of text and avoid that the model is repeating verbatim from the summary or title, we set the presence penalty to 2, the frequency penalty to 2, and the temperature to 1. For the generated texts, we target a length of 90-100 words for German and English, respectively, and 130-140 characters for Chinese.   
After generation, we use a set of heuristics for syntactic post-processing (e.g., removing unnecessary white space, adding missing punctuation, or unify the formatting of time and dates). Critical, none of these heuristics change the semantics of the generated text.
In total, we collected 30 news articles in each country (10 per category). For each of these articles, we computed up to five fakes as described above. From the resulting corpus of articles, we randomly selected 15 real and fake texts such that these are balanced across categories and article length. 

\subsection{Questionnaire}
\label{sec:method:questionnaire}
We implemented a custom framework for collecting the responses from participants with regard to the specific requirements for displaying text, image, and audio. Appendix~\ref{app:survey} shows an example view from this survey.
During the study, participants were randomly assigned to one of three media types (image, audio, and text).
After answering standard demographic questions---such as age, education, and gender---participants were asked about their knowledge on \textit{deepfakes} with a closed question (5-point Likert scale). In the next segment, we showed samples of human- and machine-generate media in a per-participant randomized order, and the participants were asked to rate how believable they were on a scale from $-3$ (definitely non-human) to $+3$ (definitely human).
Each participant saw 50\% real and 50\% fake media. This ratio was not disclosed to the participants. Participants were only informed that the data contained human- and machine-generated data. 
We showed the participants 15 human- and 15 machine-generated audio or text samples.
For images, we showed a split of 16/16 since we balanced the dataset for ethnicity and gender (\cf{} Section~\ref{sec:method:generation:images}).
After this experimental part of the questionnaire, the participants answered questions for several standard scales, which correspond to the variables of interest described in Section~\ref{sec:related:variables}: 
the \nmls{}~\citep{koc2016development} (media literacy),
the cross-cultural version of the \ahs{}~\citep{martin2022refinement} (holistic thinking), the \gts{}~\citep{yamagishi1994trust} (generalized trust), the Inglehart index~\citep{inglehart1999measuring} (political orientation), and the \crt{}~\citep{frederick2005cognitive} (cognitive reflection).
The order of the post-experiment scales were randomized per participant.
All questions were mandatory, and participants could not continue without answering all questions. 

The English and German translations of the survey were done by us. 
The Chinese version was done by our University's department of East Asian studies. 
To meet the challenges of a multilingual survey, we asked native speakers to back-translate some questions and assess the quality of the translation.

\subsection{Data Collection and Participant Compensation}
\label{sec:method:collection}

We obtained approval from our institution's Institutional Review Board (IRB), and our study protocol was deemed to comply with all ethical regulations.
In the period from June to September 2022, our online survey was distributed in the three countries (United States, Germany, and China) by the panel provider Lightspeed Research (Kantar). 
Kantar handled participant recruitment, country representative quotas, and participant compensation and has committed itself to follow the ICC/ESOMAR code of conduct~\citep{icc-07-code}.
Our panel provider did not disclose the actual participant compensation to us, and we had no influence on the compensation. They calculated a cost of 2.70€ per completed survey, which might be below the legal minimum wage in at least one of the countries surveyed. However, this compensation is in a similar range to that of crowdworking platforms~\cite{pater2021standardizing}, and is---according to our panel provider---in line with industry standards. We cannot verify this statement, as we do not have comparable data on online panelists’ compensation.

At the beginning of the study, participants were informed in detail about the purpose of the study and the use of their data.
We also informed participants that they can cancel their participation in the study at any time up until the end of the survey.
Afterwards, data withdrawal is no longer possible as participation is anonymous.
We removed any incomplete data from the dataset.
Participants gave their informed consent by checking a checkbox and clicking a button titled ``submit''.

To enhance answer quality, we implemented an attention check question that asked participants to select one specific response, \eg, ``This is an attention check please select: Neutral''.
All participants that failed our attention check were removed.
Overall, our sample consists of $n=3,002$ participants (USA: $n=1,001$; Germany: $1,000$; China: $n=1,001$).

\subsection{Data Cleaning}
\label{sec:method:cleaning}

Following~\citet{meade2012identifying}, we looked at the overall time a participant took to complete the survey and discarded every participant outside the 95\% interval.
Note that we stratified the computed interval by country and media category, \eg, we computed different intervals for China/audio, China/image, and USA/audio.
Additionally, we discarded every participant who rated every media the same, \eg, everything with ``definitely human''.
Finally,
our total number of eligible participants was $n=2,609$ ($\text{USA} = 822$; $\text{Germany} = 875$; $\text{China} = 912$).

 \section{Results}
\label{sec:analysis}

We follow our preregistration and perform an exploratory and regression analysis on the data.
We analyze the raw ratings of the participants as well as derived accuracy scores (\cf{} Section~\ref{sec:method:cleaning}).
We perform the exploratory analysis for both ratings and accuracy scores but, differing from our preregistration, only run a regression analysis on the latter.
Originally, we planned to also perform a full analysis of the ratings, but due to space constraints, we are leaving this as an open question for future work.

\subsection{RQ1: Can People Identify SOTA Generated Media?}
\label{sec:analysis:description}

\begin{table}[tp]
    \centering
    \caption{\textbf{Summary Statistics for the Dataset} We summarize the statistics of our dataset after filtering out ineligible participants (\cf{} Section~\ref{sec:method:cleaning}).
    We report aggregated age and education statistics (OECD classification).
    Ratings are on a scale from -3 (definitely non-human) to +3 (definitely human) and are centered on 0 (unsure).
\ahs{} is measured on a 7-point likert scale, while \gts{}, \fami{}, and \nmls{} are on a 5-point scale. Scores in the \crt{} can range from 0-3, and the Inglehart index (\po{}) ranges from 1-4. 
    }
    \resizebox{\linewidth}{!}{\begin{tabular}{@{}lrrrrrrrr@{}}
        \toprule
        & \multicolumn{2}{c}{USA}           &  
        & \multicolumn{2}{c}{Germany}       &  
        & \multicolumn{2}{c}{China} \\
        & \multicolumn{2}{c}{($n = 822$)}   &  
        & \multicolumn{2}{c}{($n = 875$)}   &  
        & \multicolumn{2}{c}{($n = 912$)} \\
        \midrule
        & \multicolumn{1}{c}{\textbf{n}}    & \multicolumn{1}{c}{\textbf{\%}} & 
        & \multicolumn{1}{c}{\textbf{n}}    & \multicolumn{1}{c}{\textbf{\%}} & 
        & \multicolumn{1}{c}{\textbf{n}}    & \multicolumn{1}{c}{\textbf{\%}} \\
        \multicolumn{8}{@{}l}{\textbf{Gender}} \\
        \rule{0pt}{1.8ex}\phantom{aaa}Female  &  449 &  54.62\% && 447 &  51.09\% && 433 &  47.48\% \\
        \phantom{aaa}Male    &  373 &  45.38\% && 428 &  48.91\% && 478 &  52.52\% \\
        \vspace{-1em} & & & & & & & & \\
        \multicolumn{8}{@{}l}{\textbf{Age}} \\
        \rule{0pt}{1.8ex}\phantom{aaa}18-34   &  175 &  21.29\% &&     185 &  21.19\% &&   379 &  41.60\% \\
        \phantom{aaa}35-49   &  251 &  30.54\% &&     218 &  24.97\% &&   253 &  27.77\% \\
        \phantom{aaa}50-64   &  182 &  22.14\% &&     241 &  27.61\% &&   256 &  28.10\% \\
        \phantom{aaa}65+     &  214 &  26.03\% &&     229 &  26.23\% &&    23 &   2.52\% \\
        \vspace{-1em} & & & & & & & & \\
        \multicolumn{8}{@{}l}{\textbf{Education}} \\
        \phantom{aaa}Low    &   27 &   3.28\% && 119 &  13.60\% &&   36 &   3.95\% \\
        \phantom{aaa}Medium &  349 &  42.46\% && 479 &  54.74\% &&  268 &  29.39\% \\
        \phantom{aaa}High   &  446 &  54.26\% && 277 &  31.66\% &&  608 &  66.67\% \\
        \rule{0pt}{3.5ex}&  \multicolumn{1}{c}{\textbf{mean}} & \multicolumn{1}{c}{\textbf{std}} & & \multicolumn{1}{c}{\textbf{mean}} & \multicolumn{1}{c}{\textbf{std}} & & \multicolumn{1}{c}{\textbf{mean}} & \multicolumn{1}{c}{\textbf{std}} \\
        \vspace{-0.9em} & & & & & & & & \\
        \multicolumn{8}{@{}l}{\textbf{Ratings (Human)}} \\
        \rule{0pt}{1.8ex}\phantom{aaa}Audio  &  1.60 &  1.74 && 0.99 &  1.87 && 0.33 &  1.04 \\
        \phantom{aaa}Image  &  1.65 &  1.72 && 0.97 &  1.84 && 0.42 &  0.83 \\
        \phantom{aaa}Text   &  1.70 &  1.68 && 0.89 &  1.88 && 0.73 &  0.59 \\
        \vspace{-1em} & & & & & & & & \\
        \multicolumn{8}{@{}l}{\textbf{Ratings (Machine)}} \\
        \rule{0pt}{1.8ex}\phantom{aaa}Audio  &  1.81 &  1.75 &&  0.92 &  1.97 &&  0.17 &  0.29 \\
        \phantom{aaa}Image  &  1.50 &  1.65 &&  1.10 &  1.81 &&  0.43 &  1.09 \\
        \phantom{aaa}Text   &  1.77 &  1.75 &&  0.76 &  1.89 &&  0.49 &  0.27 \\
        \vspace{-1em} & & & & & & & & \\
        \multicolumn{8}{@{}l}{\textbf{Accuracy}} \\
        \rule{0pt}{1.8ex}\phantom{aaa}Audio      &  50.57\% &  10.38\% &&  59.15\% &  13.98\% &&  51.73\% &  11.65\% \\
        \phantom{aaa}Image      &  48.58\% &  10.62\% &&  45.90\% &  11.80\% &&  49.93\% &  10.47\% \\
        \phantom{aaa}Text       &  51.50\% &   9.40\% &&  54.48\% &  10.30\% &&  52.45\% &  10.22\% \\
        \vspace{-1em} & & & & & & & & \\
\vspace{-1em} & & & & & & & & \\
        \multicolumn{8}{@{}l}{\textbf{Predictors}} \\
        \rule{0pt}{1.8ex}\phantom{aaa}CRT & 0.49 & 0.83 & & 1.04 & 1.04 & & 1.49 & 1.10 \\
        \phantom{aaa}FAM & 1.04 & 1.32 & & 1.00 & 1.21 & & 1.38 & 1.16 \\
        \phantom{aaa}PO & 1.30 & 1.03 & & 1.20 & 1.04 & & 1.12 & 0.86 \\
        \phantom{aaa}AHS & 4.73 & 0.56 & & 4.97 & 0.62 & & 4.89 & 0.49 \\
        \phantom{aaa}GTS & 3.44 & 0.74 & & 3.42 & 0.66 & & 3.81 & 0.65 \\
        \phantom{aaa}NMLS CC & 3.61 & 0.58 & & 3.65 & 0.57 & & 3.83 & 0.49 \\
        \phantom{aaa}NMLS CP & 2.69 & 1.06 & & 2.17 & 0.96 & & 3.57 & 0.70 \\
        \phantom{aaa}NMLS FC & 3.69 & 0.63 & & 3.74 & 0.56 & & 3.84 & 0.52 \\
        \phantom{aaa}NMLS FP & 3.01 & 1.04 & & 2.72 & 0.91 & & 3.80 & 0.60 \\

        \bottomrule
    \end{tabular}
    }
    \label{tab:summary}
\end{table}

 \begin{table}[t]
    \centering
    \caption{\textbf{Percentage of Samples Rated as Human-Rated}
    We report the percentage of samples rated as human-generated per country and media type, where we aggregate the human-generated ratings ($+1, \,+2, \,+3$).
    }
    \resizebox{\linewidth}{!}{\begin{tabular}{@{}lrrrrrrrr@{}}
\rule{0pt}{2.4ex}&  \multicolumn{2}{c}{USA}  
         && \multicolumn{2}{c}{Germany} 
         && \multicolumn{2}{c}{China}\\
         \rule{0pt}{2.4ex}&  \multicolumn{1}{c}{Human} & \multicolumn{1}{c}{Machine} 
         && \multicolumn{1}{c}{Human} & \multicolumn{1}{c}{Machine} 
         && \multicolumn{1}{c}{Human} & \multicolumn{1}{c}{Machine} \\
         \midrule
        Audio & 73.67\% & 72.20\% && 76.91\% & 58.93\% && 58.14\% & 54.93\% \\
        Image & 74.23\% & 77.70\% && 70.86\% & 78.64\% && 61.14\% & 61.17\% \\
        Text  & 72.70\% & 70.28\% && 67.52\% & 59.27\% && 67.88\% & 62.75\% \\
    \end{tabular}
    }
    \label{tab:rating:overview}
    \vspace{-0.5em}
    \begin{flushleft}
    \scriptsize{\% Rated as Human}\\
    \end{flushleft}
\end{table} 

We report a summary of our dataset in Table~\ref{tab:summary}. 
Generally speaking, participants in all countries and across all media types predominantly rated media human-generated, regardless of whether the underlying media was actually created by a human or a machine.
The average rating peaks at $1.81$ for audio in the US and bottoms out at $0.17$ for audio in China.
Kruskal-Wallis one-way ANOVA showed a statistically significant difference across all media for both human-generated (audio: $H(2) = 339.80$ $p < 0.001$; image: $H(2) = 142.60$ $p < 0.001$; text: $H(2) = 87.19$ $p < 0.001$) and machine-generated media (audio: $H(2) = 331.56$ $p < 0.001$; image: $H(2) = 257.55$ $p < 0.001$; text: $H(2) = 143.99$ $p < 0.001$).
We perform post-hoc Mann-Whitney-U tests (Bonferroni corrected) and report the results in Table~\ref{tab:rating:diff}.
We find statistically significant differences between all pairs,  except for USA-Germany (Human-Audio), Germany-China (Machine-Audio), USA-Germany (Machine-Image), and USA-China (Human-Text).

\begin{table*}[t]
    \caption{\textbf{Pair-wise Mann-Whitney U Tests Between the Country Ratings} We report the difference in mean ratings between different country pairs.
    {\small * $p < 0.05$; ** $p < 0.01$; *** $p < 0.001$ - Bonferroni equivalent}
    }
    \textbf{Human-Generated}
    \vfill
    \vspace{1em}
    \vfill
    \begin{subtable}[h]{.32\textwidth}
        \centering
        \begin{tabular}{@{}rrr@{}}
            & Germany & China \\
            \midrule
            USA & 0.03\,±\,2.26* & 0.61\,±\,2.41*** \\
            Germany & N/A\phantom{***} & 0.62\,±\,2.31*** \\
        \end{tabular}
\label{tab:rating:diff:audio:real}
    \end{subtable}
    \hfill
    \begin{subtable}[h]{.32\textwidth}
        \centering
        \begin{tabular}{@{}rrr@{}}
            & Germany & China \\
            \midrule
            USA & 0.15\,±\,2.18*** & 0.45\,±\,2.35*** \\
            Germany & N/A\phantom{***} & 0.31\,±\,2.35*** \\
        \end{tabular}
\label{tab:rating:diff:image:real}
    \end{subtable}
    \hfill
    \begin{subtable}[h]{.32\textwidth}
        \centering
        \begin{tabular}{@{}rrr@{}}
             & Germany & China \\
             \midrule
            USA & 0.27\,±\,2.17*** & 0.09\,±\,2.36\phantom{***} \\
            Germany & N/A\phantom{***} & -0.19\,±\,2.35*** \\
        \end{tabular}
\label{tab:rating:diff:text:real}
    \end{subtable}
    \vfill
    \vspace{1.5em}
    \vfill
\textbf{Machine-Generated}
    \vfill
    \vspace{1em}
    \vfill
    \begin{subtable}[h]{.32\textwidth}
        \centering
        \begin{tabular}{@{}rrr@{}}
            & Germany & China \\
            \midrule
            USA & 0.57\,±\,2.34*** & 0.65\,±\,2.49*** \\
            Germany & N/A\phantom{***} & 0.10\,±\,2.47\phantom{***} \\
        \end{tabular}
\begin{center}
            \textbf{Audio}
        \end{center}
        \label{tab:rating:diff:audio:fake}
    \end{subtable}
    \hfill
    \begin{subtable}[h]{.32\textwidth}
        \centering
        \begin{tabular}{@{}rrr@{}}
            & Germany & China \\
            \midrule
            USA & 0.03\,±\,2.12\phantom{***} & 0.56\,±\,2.35*** \\
            Germany & N/A\phantom{***} & 0.53\,±\,2.24*** \\
        \end{tabular}
\begin{center}
            \textbf{Image}
        \end{center}
        \label{tab:rating:diff:image:fake}
    \end{subtable}
    \hfill
    \begin{subtable}[h]{.32\textwidth}
        \centering
        \begin{tabular}{@{}rrr@{}}
            & Germany & China \\
            \midrule
            USA & 0.40\,±\,2.23*** & 0.18\,±\,2.42*** \\
            Germany & N/A\phantom{***} & -0.23\,±\,2.37*** \\
        \end{tabular}
\begin{center}
            \textbf{Text}
        \end{center}
        \label{tab:rating:diff:text:fake}
    \end{subtable}
    \vfill
    \label{tab:rating:diff}
\end{table*}

In Table~\ref{tab:rating:overview}, we investigate these differences in more detail and report the percentage of human-rated samples per country and media type.
In general, all media, across all categories, were predominantly rated as human-generated, even if they were artificially created.
US participants show a larger tendency to predict a sample as being human-generated (roughly between 70-78\% of the samples) independent of the category.
German (between 58-76\%) and Chinese (between 54-68\%) participants exhibit a wider range of ratings but still predominantly rate the samples as human-generated.
Looking at the results in more detail, machine-generated audio data is more often detected by German participants compared to participants from the US.
While Chinese participants detect even more machine-generated data, they also more often interpret human-generated data as artificial, which leads to an overall smaller accuracy (see Table~\ref{tab:summary}).

Interestingly, US as well as German participants have a tendency to rate machine-generated images more often as human-generated than real images, which does not hold for Chinese participants. In the case of text data, US participants are more often convinced that artificial texts are written by a human than Chinese participants. German participants are even slightly better in identifying machine-generated text. This might be due to lower quality of the generated text due to fewer training samples of the model.

\begin{figure}[t]
    \centering
    \includegraphics[width=.98\linewidth]{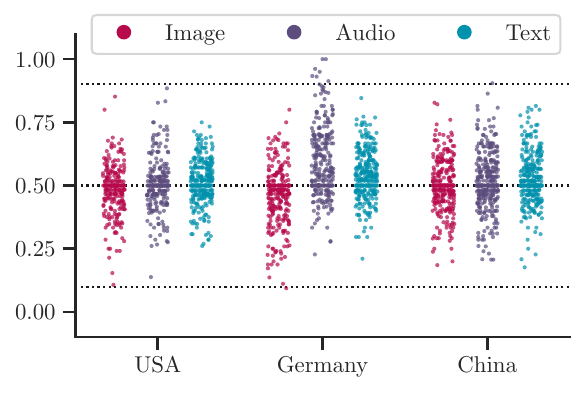}
    \caption{
        \textbf{Accuracy of the Participants} We plot the accuracy of our participants correctly identifying a media as fake or real.
        The levels indicate $10\%$, $50\%$ and $90\%$ accuracy.
    }
    \label{fig:acc:overview}
\end{figure} \begin{figure*}[t]
    \centering
    \includegraphics[width=.98\linewidth]{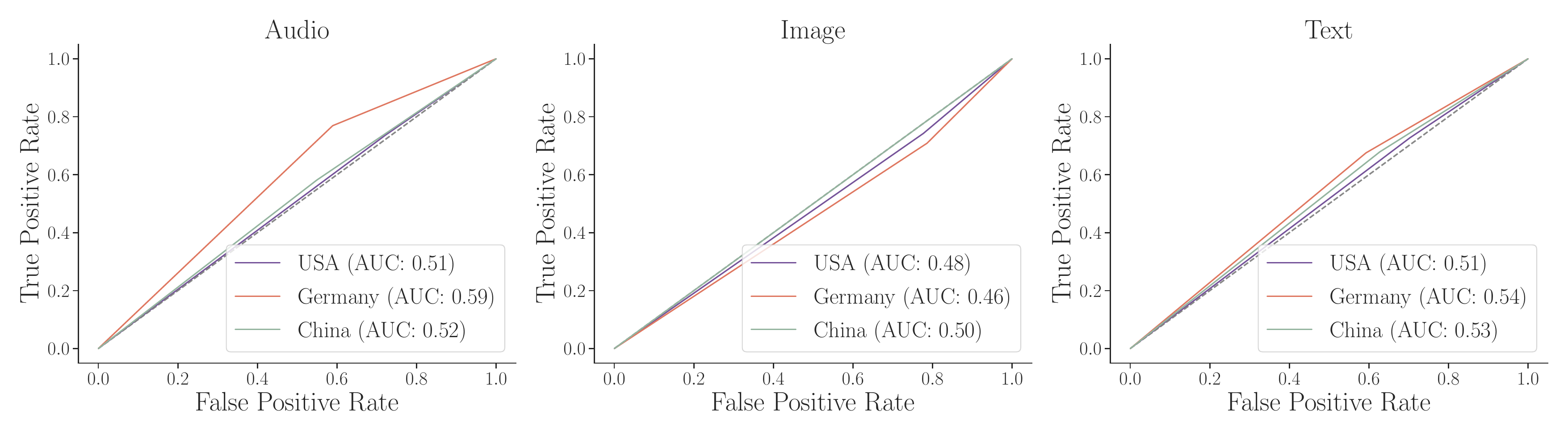}
    \caption{
        \textbf{ROC-Curve for Different Media Types and Country Pairs} We plot the Receiver Operating Characteristic (ROC) across media type and country pairs.
        We also report the Area-Under-the-Curve (AUC) values.
    }
    \label{fig:roc:overview}
\end{figure*} \begin{table*}[t]
    \caption{\textbf{Pair-wise Mann-Whitney U Tests Between the Country Accuracies} We report the difference in accuracy between different country pairs.
    {\small * $p < 0.05$; ** $p < 0.01$; *** $p < 0.01$ - Bonferroni equivalent}
    }
    \begin{subtable}[h]{.32\textwidth}
        \centering
        \begin{tabular}{@{}rrr@{}}
             & Germany & China \\
                \midrule
            USA & -8.49\,±\,17.69*** & -0.40\,±\,15.43\phantom{***} \\
            Germany & N/A\phantom{{***}} & 8.15\,±\,17.72*** \\
        \end{tabular}
        \begin{center}
            \textbf{Audio}
        \end{center}
        \label{tab:acc:diff:audio}
    \end{subtable}
    \hfill
    \begin{subtable}[h]{.32\textwidth}
        \centering
        \begin{tabular}{@{}rrr@{}}
            & Germany & China \\
            \midrule
            USA & 2.35\,±\,14.80*\phantom{**} & -1.37\,±\,14.83\phantom{***} \\
            Germany & N/A\phantom{{***}} & -4.06\,±\,16.23*** \\
        \end{tabular}
        \begin{center}
            \textbf{Image}
        \end{center}
        \label{tab:acc:diff:image}
    \end{subtable}
    \hfill
    \begin{subtable}[h]{.32\textwidth}
        \centering
        \begin{tabular}{@{}rrr@{}}
            & Germany & China \\
            \midrule
            USA & -2.58\,±\,13.60**\phantom{*} & -1.17\,±\,13.28\phantom{***} \\
            Germany & N/A\phantom{{***}} & 1.49\,±\,14.25*\phantom{**} \\
        \end{tabular}
        \begin{center}
            \textbf{Text}
        \end{center}
        \label{tab:acc:diff:text}
    \end{subtable}
    \label{tab:acc:diff}
\end{table*} \begin{figure*}[t]
     \centering
     \includegraphics[width=.98\linewidth]{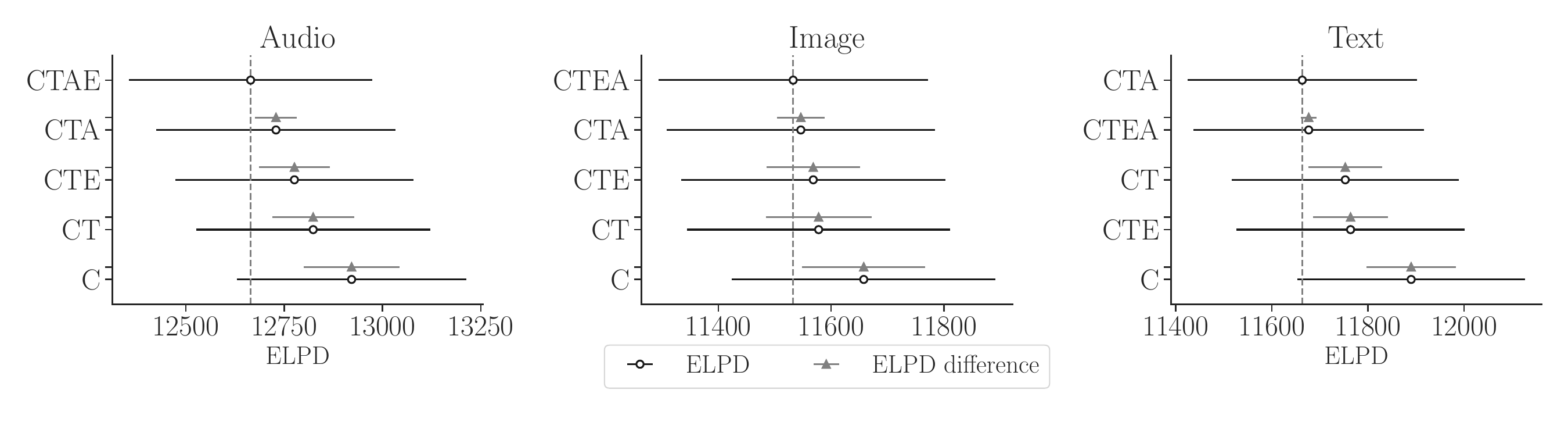}
\caption{\textbf{Comparing the ELPD of different model iterations} 
    We plot the ELPD of each model, as well as the expected difference for each model w.r.t.\ the best model.
    Lower ELPD values indicate a better model fit to the data.
    Note that a specific ELPD value is meaningless, it derives meaning from comparison with the other models' values.
    The figure is ordered from top to bottom, with the best performing model being on top.
    \newline
    {\small The models are coded using the variables included: (C)ountry, Median (T)ime, (A)ge, (E)ducation.}
}
    \label{fig:model_comparison}
\end{figure*}
 
We also analyze the overall accuracy of our participants in Figure~\ref{fig:acc:overview}.
We observe the biggest difference between the groups in the audio condition (USA $50.57\pm10.38$; Germany $59.15\pm13.98$; China $51.73\pm11.65$), where Germans perform better than US or Chinese participants.
The difference is less pronounced for images, where Germany performs the worst (USA $48.58\pm9.40$; Germany $45.90\pm11.80$; China $49.93\pm10.47$).
The accuracy for text is most similar between the three groups (USA $51.50\pm9.40$; Germany $54.48\pm10.30$; China $52.45\pm10.22$).
Kruskal-Wallis one-way ANOVA showed a statistically significant ($p < 0.001$) difference for the accuracy on audio and image, but not on text ($p = 0.005$).
We also run post-hoc paired Mann-Whitney U tests.
The results are shown in Table~\ref{tab:acc:diff}.
Germany differs for audio and image, while all countries are fairly close for text.

Finally, to better visualize our overall results, we also plot the \emph{Receiver Operating Characteristic} (ROC) curve in Figure~\ref{fig:roc:overview}.
This curve visualizes the \emph{True Positive Rate} (TPR) against the \emph{False Positive Rate} (FPR) of participant ratings.
A TPR of $1.0$ and FPR of $0.0$ would indicate that every media is rated correctly.
Again, we can observe that participants are mostly guessing with the exception of German audio data, which is a clear outlier.

We hypothesize that this might be due to a lower quality of the German audio data.
From anecdotal evidence, the German audio samples sound more ``robotic'' and noisier than samples in other languages.
Based on the data we have collected, we cannot make a full conclusion about the cause.
However, note that the survey prompts are almost identical for the different media types, so errors in the experimental setup would have shown up in the other media types as well.

\begin{insightbox}
\finding{
Across all media types and countries, we observe that artificially generated samples are almost indistinguishable from ``real'' media.
Participants predominantly rated artificially generated media as human-generated and performed even worse than random guessing for images.
Surprisingly, all countries are quite close to each other w.r.t.\ performance.
This is surprising, as English media is often believed to be far ahead of media in other languages~\citep{rae2021scaling}.
An exception is the German audio data, where the participants perform significantly better.
}
\end{insightbox}

\subsection{Modelling Choices}
\label{sec:analysis:acc:model}

We now analyze the accuracy of our participants for human- and machine-generated media separately by running a regression analysis to analyze the probability of correctly labeling a given media.
Following established guidelines for analyzing cross-cultural data~\citep{deffner2021causal}, we use a Bayesian multilevel-binomial linear mixed effect model predicting the probability of correctly classifying human- or machine-generated media.
For our analysis, we choose a Bayesian framework to obtain our estimates instead of a Frequentist one.
Similar models can be built with maximum likelihood estimation~\citep{goldstein2011multilevel}, but it has been shown that Bayesian methods produce more stable estimates~\citep{gelman2006data,rabe2006multilevel,stegmueller2013many}.

Compared to the Frequentist approach, which treats model parameters as point estimates (averages), the Bayesian regression treats parameters as random variables (probability distributions).
More specifically, when we use an ordinary least square regression, we want to estimate the most likely parameters; in a Bayesian regression analysis, we are instead interested in estimating a range of likely values for this parameter.
This estimated distribution is referred to as the \emph{posterior distribution}.

We choose a multilevel (or hierarchical) model to account for the different subgroups (i.e., demographic variables). 
Furthermore, we allow the estimates to vary at the country level, while also letting these results influence the population estimate.
For example, the estimate for US participants in a particular age group may inform the estimate for German participants in a similar age group.
An additional benefit of a Bayesian approach is that we can model locality, e.g., we can model that age groups that are closer in age are more correlated.

We first examine a comparison between different countries in Section~\ref{sec:analysis:acc:realfake} and choose our model to explain the differences between countries.
We then build on that analysis and assess the influence of predictor values independently of the culture (i.e., on a population level) in Section~\ref{sec:analysis:acc:personal}.

\begin{figure}[t]
     \centering
     \begin{subfigure}[b]{\linewidth}
         \centering
         \includegraphics[width=\linewidth,trim={0 20pt 0 10pt},clip]{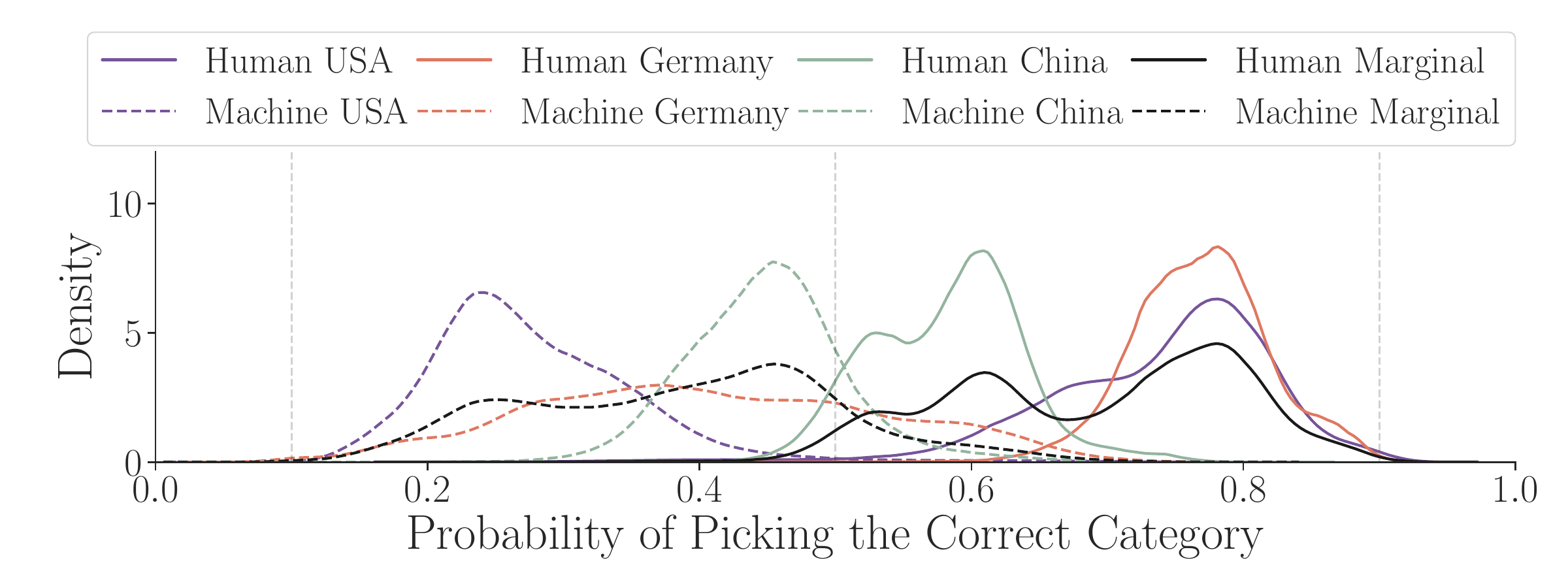}
         \caption{Audio}
        \label{fig:acc:posterior:audio}
     \end{subfigure}
     \vfill
     \begin{subfigure}[b]{\linewidth}
         \centering
         \includegraphics[width=\linewidth,trim={0 20pt 0 10pt},clip]{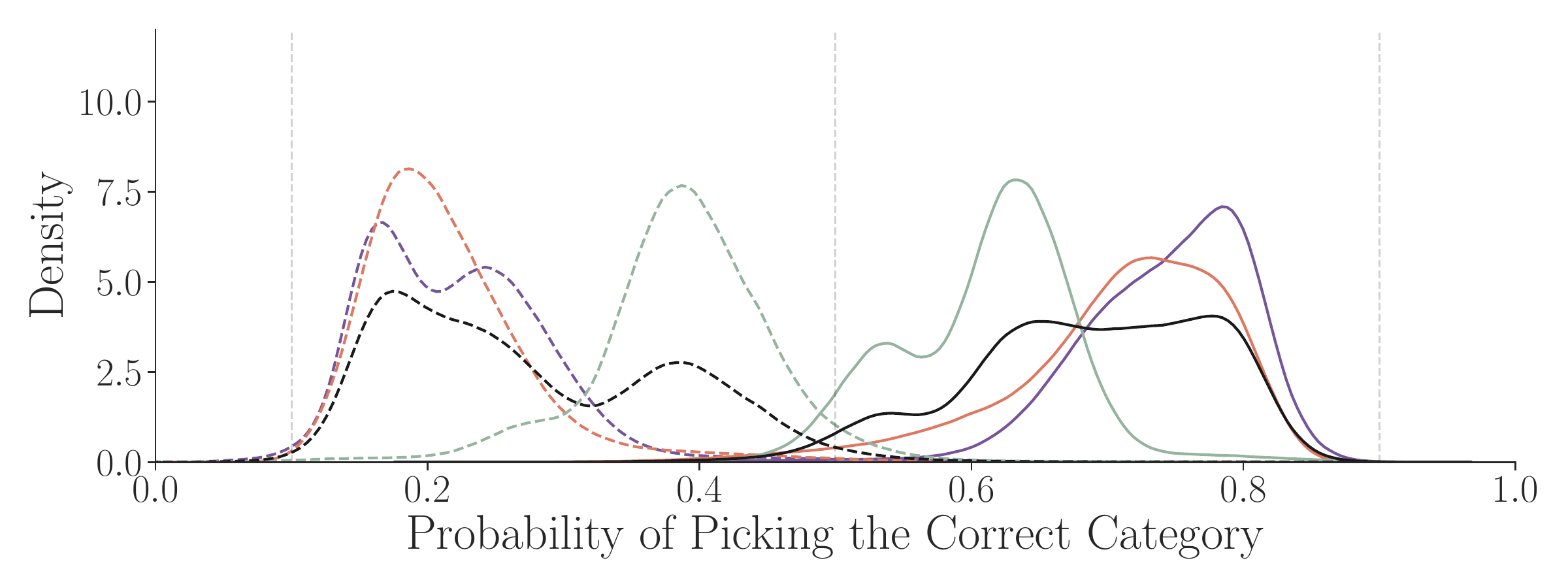}
         \caption{Image}
        \label{fig:acc:posterior:image}
     \end{subfigure}
     \vfill
     \begin{subfigure}[b]{\linewidth}
         \centering
         \includegraphics[width=\linewidth,trim={0 20pt 0 10pt},clip]{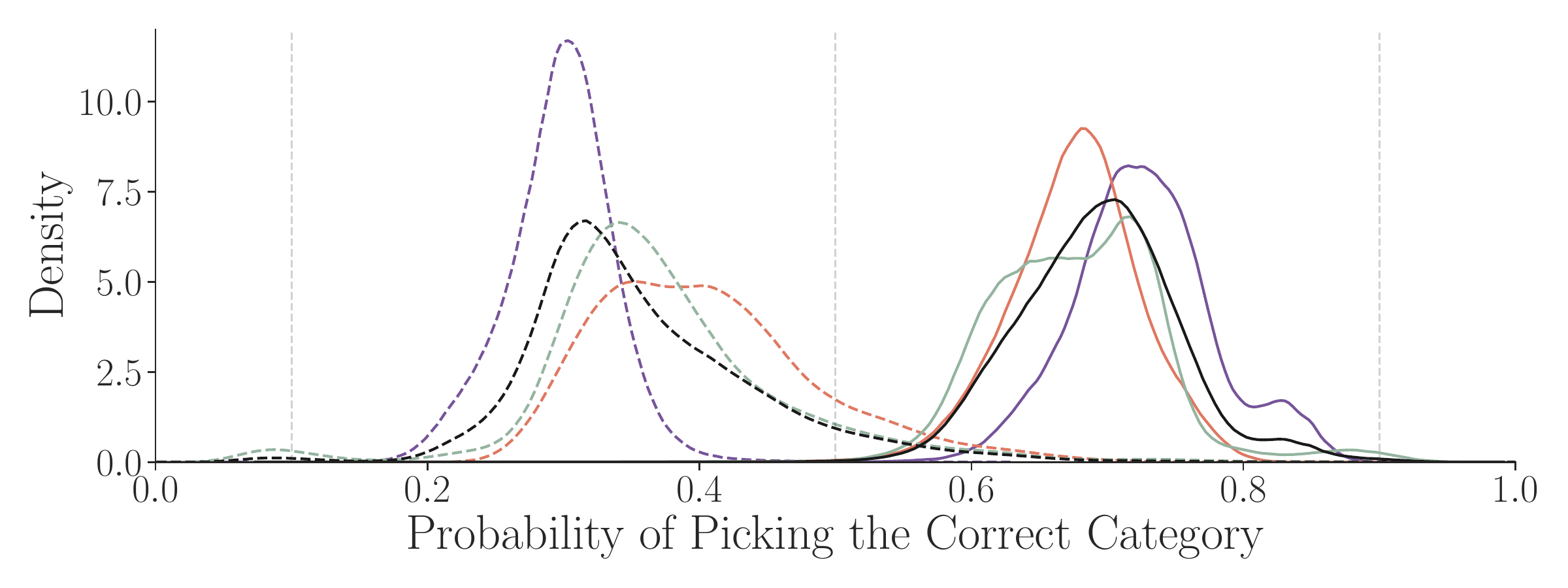}
         \caption{Text}
        \label{fig:acc:posterior:text}
     \end{subfigure}
    \caption{\textbf{Predicted Posterior Probability of the Regression Models} We plot the predicted posterior probability of picking the correct results by media type. 
    We plot the results separated by country as well as the marginal 
    over all countries.
    The levels indicate $0.1$, $0.5$, and $0.9$.
    }
    \vspace{-1.5em}
    \label{fig:acc:posterior}
\end{figure} \subsection{RQ2: Which Demographic Factors Influence the Identification Accuracy?}
\label{sec:analysis:acc:realfake}

We start by analyzing the influence of different demographic variables across cultures.
To that end, we compare several model iterations for each media type, where each iteration adds specific demographic variables.
The full model includes varying intercepts per country and three correction terms: i) one modelling the age range of a participant, ii) one modelling their education level, and iii) a term for the median time taken per stimulus.
We partially pool these variables to obtain more stable estimates.
A full model description and further details can be found in Appendix~\ref{app:model}.

To compare our models, we use the \emph{Expected Log Pointwise Predictive Density} (ELPD), which is a measure  on how well a given model generalizes to new data.
We compute this metric by leave-one-out cross validation; that is, we fit a model on all but one data point, compute the log predictive densities for the left out point (i.e., measure how well the model predicts the left out point), and repeat that process for each data point.
In practice, we do not refit the model each time, instead there are reliable ways to estimate this process~\cite{vehtari2017practical}.

We plot the results in Figure~\ref{fig:model_comparison}, where a smaller ELPD value indicates better model fit to the data.
Additionally, we also report the expected pointwise difference for each model w.r.t.\ the best model found.
The smaller the expected difference between the models' ELPD, the more similar the predictions.
Note that we also tried model variants omitting the time correction term, but they performed worse compared to the rest.
For brevity, we omit them from the plot.

Audio and image data are best explained by the full model (CTAE).
The difference is more pronounced for audio data, where including an education term is a clear advantage over the model that only includes the age term.
We analyze these differences more closely in Section~\ref{sec:analysis:acc:realfake:demographic}.
However, while the contrast is more stark for audio, the difference between the image models is closer.
Finally, when we look at text data, the best performing model only considered median time and age of a participant.

\subsubsection{Overall Results}
\label{sec:analysis:acc:realfake:overall}

In Figure~\ref{fig:acc:posterior},  we plot the predicted posterior distribution of the probability of picking the correct category, both as the marginal distribution (integrated over all data) and split into the different conditional distributions given the respective countries.
Note that we discuss the posterior probability predicted by our model for our sample, not the exact probabilities implied by our sample.
The difference might be subtle but is important.
When we discuss the exact statistics of our sample, we would assume that we have accurately sampled the underlying population, which is rarely true in reality (\cf{} Section~\ref{sec:limitations}).
Instead we build a statistical model accounting for the uncertainty implied by our sampling method, which in turn allows us to better approximate the (real) underlying population~\citep{mcelreath2020statistical,deffner2021causal,vehtari2017practical}.

These results give further insights into our initial rating observations presented in Section~\ref{sec:analysis:description}.
We can again observe the trend that people are fairly good at identifying real media and are worse when it comes to identifying fakes.
For example, machine-generated audio data is easier to identify ($16.83\%$ of the marginal probability mass above $.5$) than both image ($1.28\%$) and text data ($7.83\%$).
This trend even persists when we discount German audio data ($10.56\%$).
The distributions also give a better perception of the variability of the participants.
For example, the density for machine-generated German audio data is fairly spread out, encompassing individuals who are really good ($27.39\%$ of the probability mass above $.5$) and really bad ($13\%$ lower than $.25$) at identifying generated data.

\begin{insightbox}
\finding{
In our preregistration, we hypothesized that image data would be the most convincing, audio to be in the middle, and text to be the least convincing.
In contrast, our results suggest that fake audio data is the least convincing.
We conjecture two plausible explanations: First, as introduced in Section~\ref{sec:method:generation}, we generate audio data by using a text-to-speech pipeline, i.e., we supply a textual representation of the audio we want to generate.
For text data, we supply the headline together with a short summary, which the model expands to a full paragraph.
In hindsight, the text model has many different but equally valid ways to generate the paragraph. The audio model, on the other hand, has to both match the exact text and generate believable human voices with matching pitch and rhythm.
Second, we employ the most advanced version of OpenAI's GPT-3 models, which dwarf our speech model in both size and the volume of training data~\citep{brown-20-language,kong2020hifi}.
}
\end{insightbox}

\subsubsection{Paired Comparisons}
\label{sec:analysis:acc:realfake:paried}

We also performed a paired comparison between countries. To this end, we compute the contrast distributions between different country pairs, as looking at the posterior distributions directly can be deceiving~\citep{mcelreath2020statistical}.
To compute the contrast distributions, we randomly draw 20,000 pairs of samples from the distributions we want to compare and subtract those samples from each other to obtain a ``distribution of differences''.

\begin{figure}[t!bp]
     \centering
     \begin{subfigure}[b]{\linewidth}
        \centering
        \includegraphics[width=\linewidth,trim={0 20pt 0 40pt},clip]{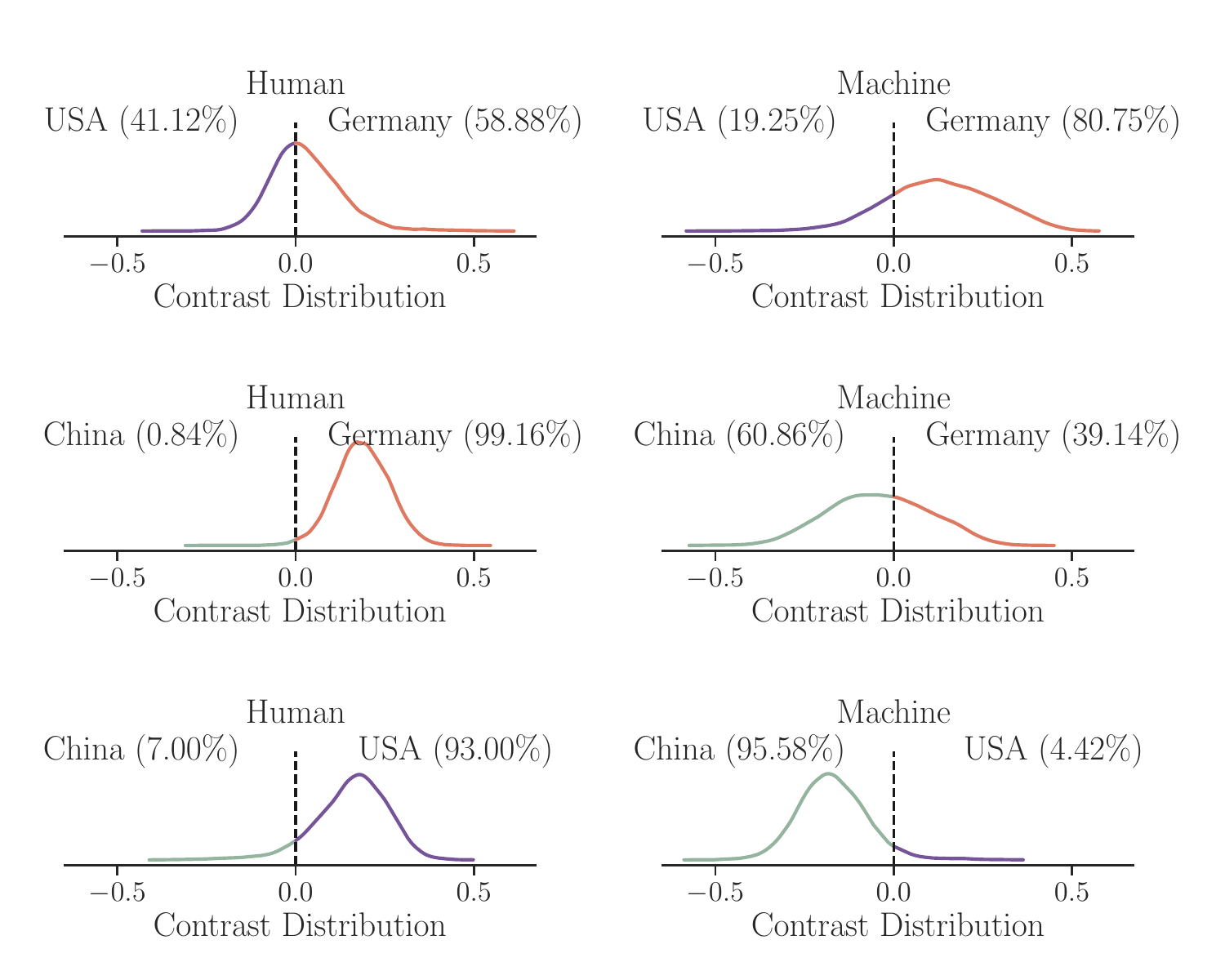}
        \caption{Audio}
        \label{fig:acc:contrast:audio}
     \end{subfigure}
     \vfill
     \begin{subfigure}[b]{\linewidth}
        \centering
        \includegraphics[width=\linewidth,trim={0 20pt 0 10pt},clip]{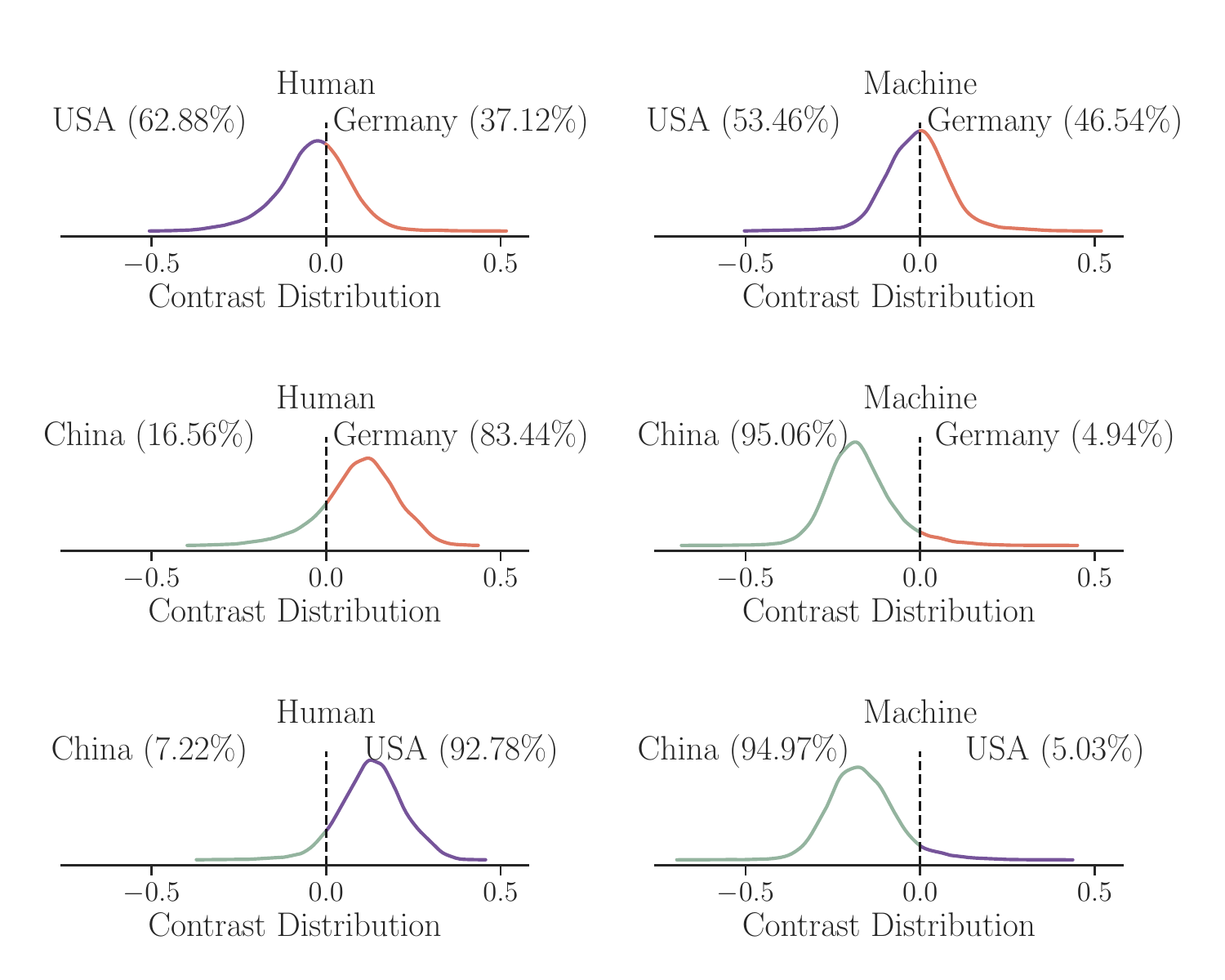}
        \caption{Image}
        \label{fig:acc:contrast:image}
     \end{subfigure}
     \vfill
     \begin{subfigure}[b]{\linewidth}
        \centering
        \includegraphics[width=\linewidth,trim={0 20pt 0 10pt},clip]{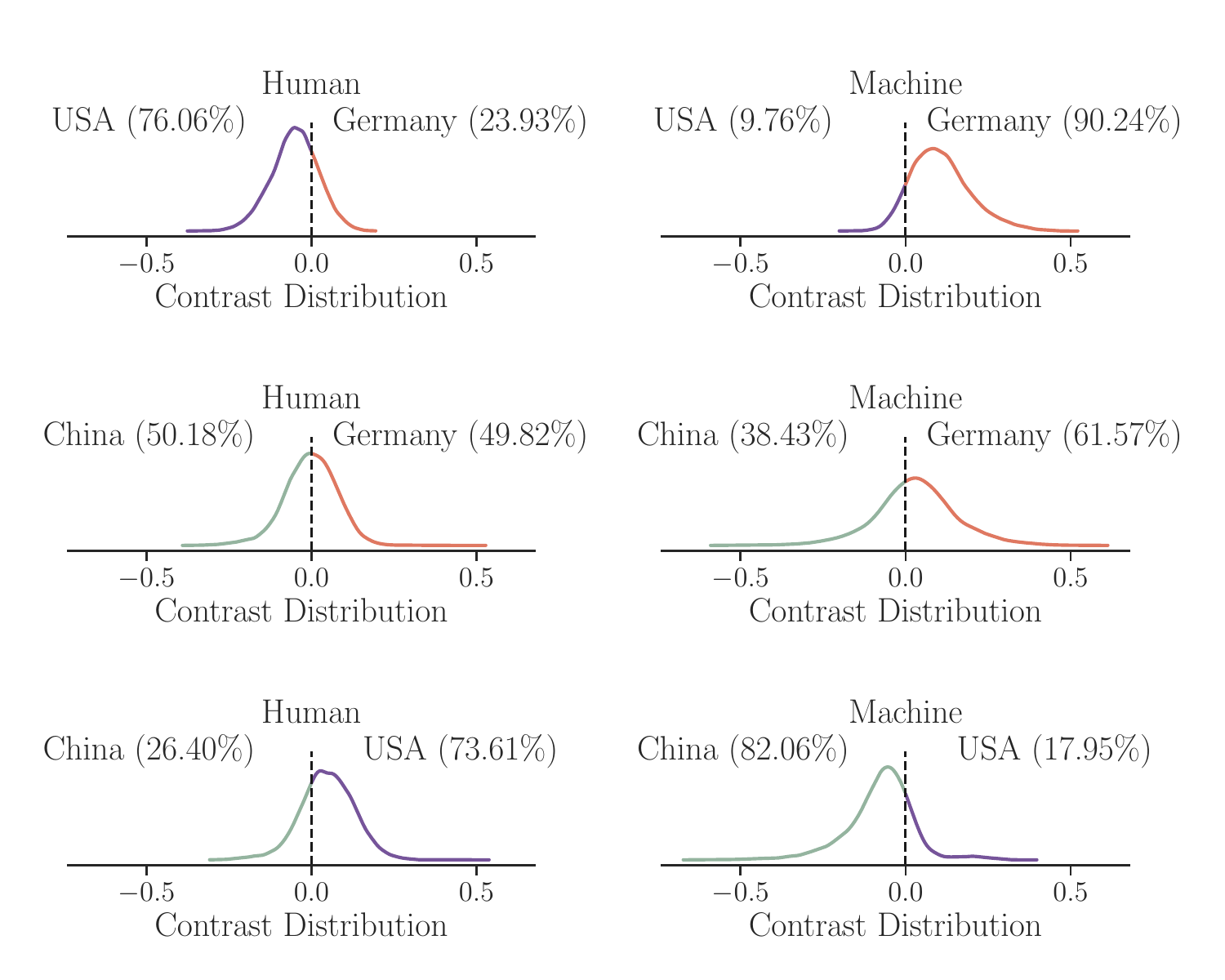}
        \caption{Text}
        \label{fig:acc:contrast:text}
     \end{subfigure}
    \caption{\textbf{Contrast Distributions Comparing Different Countries} We plot the contrast distribution obtained by subtracting one country's posterior distribution from the other. 
    }
    \label{fig:acc:contrast}
\end{figure}

For example, if we want to compare participants from China and Germany for artificially generated audio, we would subtract each sample from the Chinese posterior distribution from a sample from the German posterior distribution.
If German participants are generally better at identifying machine-generated media, the majority of the distribution would consist of positive values.
The resulting contrast distributions are shown in Figure~\ref{fig:acc:contrast}.
Indeed, the contrast distribution between Germany and China for machine-generated audio samples is slightly skewed towards Chinese participants (\cf{} Figure~\ref{fig:acc:contrast:audio}).
That indicates that Chinese participants might have a small edge over German participants.
This is not immediately obvious from Figure~\ref{fig:acc:posterior:audio}, since German participants exhibit a fairly high variance, resulting in both people who are very good and people who are very bad at detecting machine-generated samples.

\textbf{Audio:}  We start by examining audio data by inspecting the respective contrast distributions in Figure~\ref{fig:acc:contrast:audio}.
First we compare participants from Germany against participants from the USA: German participants are slightly better compared to those from the USA when detecting human-generated audio data (for $60.95\%$ of the sampled pairs the probability of picking the correct label was higher for the German samples than for the US samples) and are clearly better at recognizing machine-generated audio samples ($79.80\%$).
As stated before, this may be explained by the lower quality of German machine-generated audio.
When compared to Chinese participants, the German participants are clearly better at recognizing human-generated data ($99.01\%$) but perform slightly worse on machine-generated audio samples ($39.34\%$).
Finally, we compare participants from the USA and China, showing that US participants are generally better at identifying human-generated samples and worse at identifying machine-generated ones.
These results agree with our initial observations about the ratings:
Chinese participants show a lower probability of rating a sample as human-generated, no matter if it stems from a human or a machine.
While US and Chinese participants effectively randomly guess their choices, they do it at different rates, reflected by the fact that they outperform the other in different categories (while being equal overall).
German participants on the other hand clearly outperform the other regions in one category while being about equal in the other.

\textbf{Image:} When comparing images, US participants perform slightly better compared to Germans on both human- ($62.23\%$ favors USA) and machine-generated samples ($54.53\%$).
The difference is bigger when comparing Germany and China, where Chinese participants are clearly better in recognizing machine-generated content ($94.93\%$ favors China), the ratio is closer for human-generated content (only $83.15\%$ favors Germany).
Finally, we can again find a pattern where the US and Chinese participants are distinctively better at either category.
These results mirror our initial findings, where Chinese and German participants (Germans being worse) clearly differ, but the rest of the pairs are more closer together.

\textbf{Text:} For text data, we can confirm our previous observation that it yields the most similar results (\cf{} Section~\ref{sec:analysis:description}), with no clear standout.
Germany and the US are the most different, with German participants outperforming the US participants more clearly on machine-generated media ($86.78\%$ of the distribution favors Germany) than the US participants do on human-generated samples ($74.17\%$).
The difference between Chinese and German participants is much smaller ($59.20\%$ and $49.80\%$, respectively), while the contrast distributions for the USA and China are about equal but slightly favor Chinese participants.

\begin{insightbox}
\finding{
Our paired comparisons explain the phenomenon observed in Section~\ref{sec:analysis:description} where countries often perform similarly overall, even though they perform significantly differently when compared on a per-category basis (human-/machine-generated).
Examining the contrast distributions, we observe that they often seem like mirror images of each other.
In other words, while one country performs better on machine-generated samples, the paired country performs better on human-generated samples.
This, in-turn, leads to a similar overall accuracy.
This is most noticeable for image and text and, when comparing USA and China, for audio as well.
}
\end{insightbox}

\subsubsection{Demographic Variables}
\label{sec:analysis:acc:realfake:demographic}
\begin{figure}[t!bp]
     \centering
     \begin{subfigure}[b]{\linewidth}
         \centering
         \includegraphics[width=\linewidth]{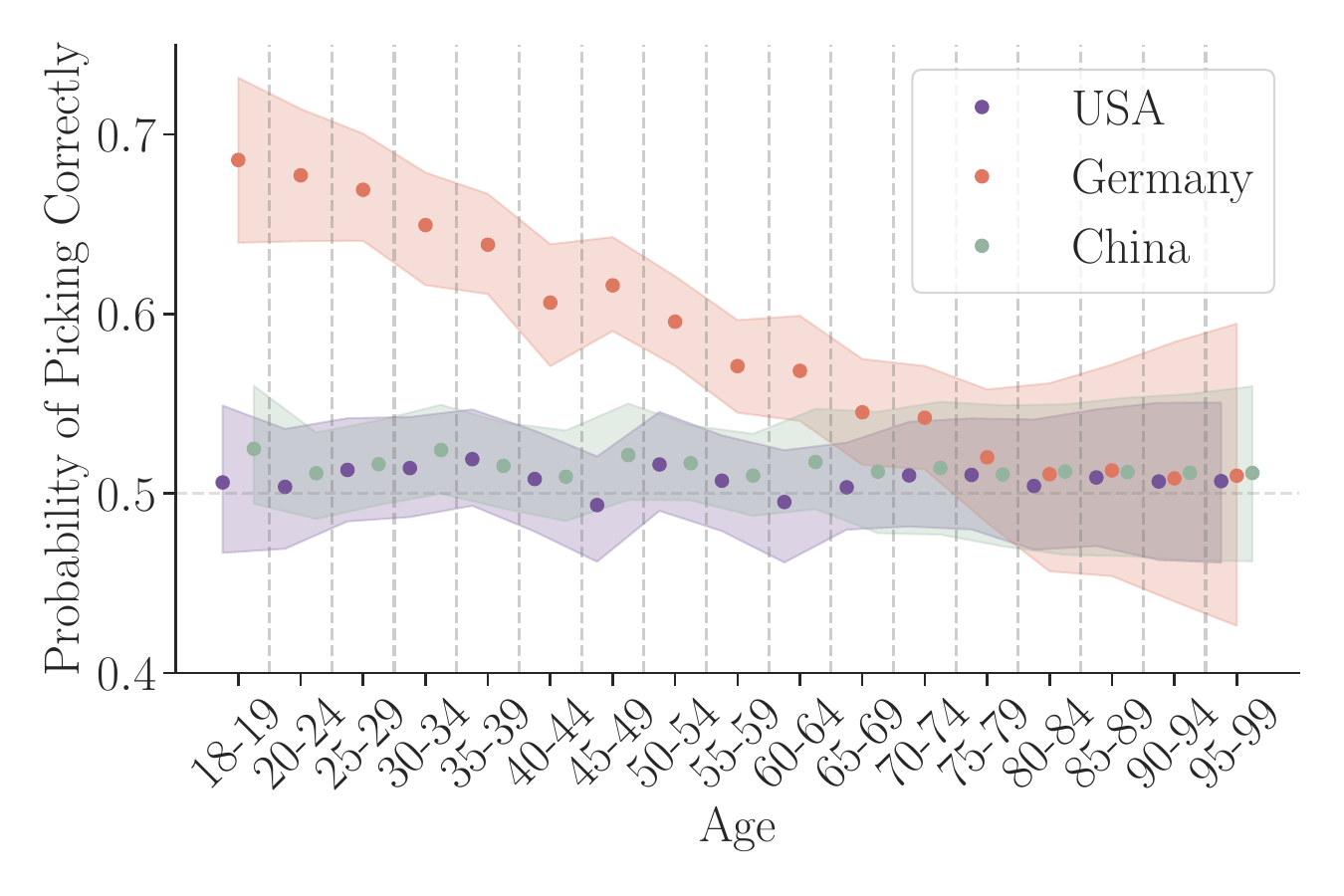}
         \caption{Audio}
        \label{fig:demographic:age:audio}
     \end{subfigure}
     \vfill
     \begin{subfigure}[b]{\linewidth}
         \centering
         \includegraphics[width=\linewidth]{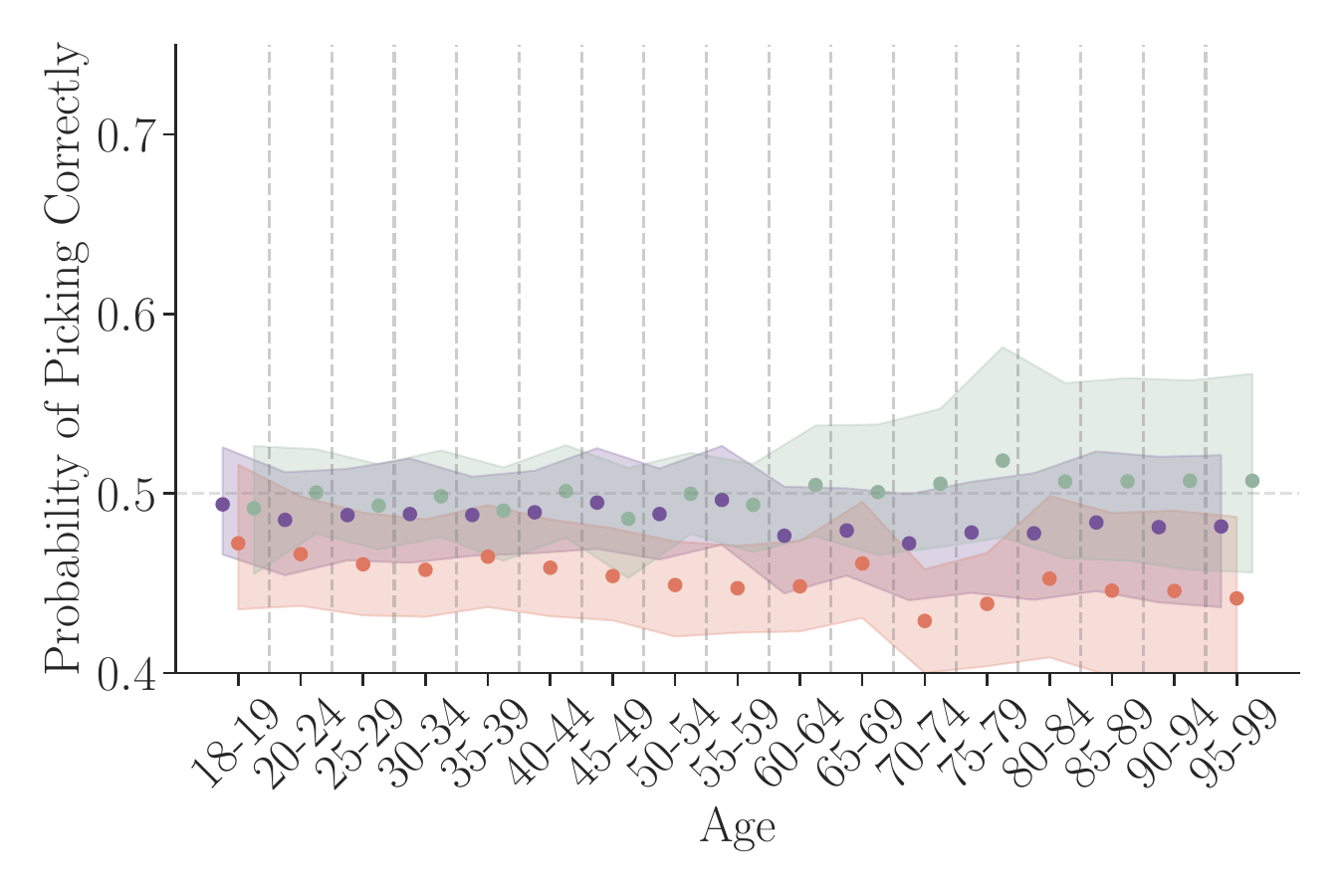}
         \caption{Image}
        \label{fig:demographic:age:image}
     \end{subfigure}
     \vfill
     \begin{subfigure}[b]{\linewidth}
         \centering
         \includegraphics[width=\linewidth]{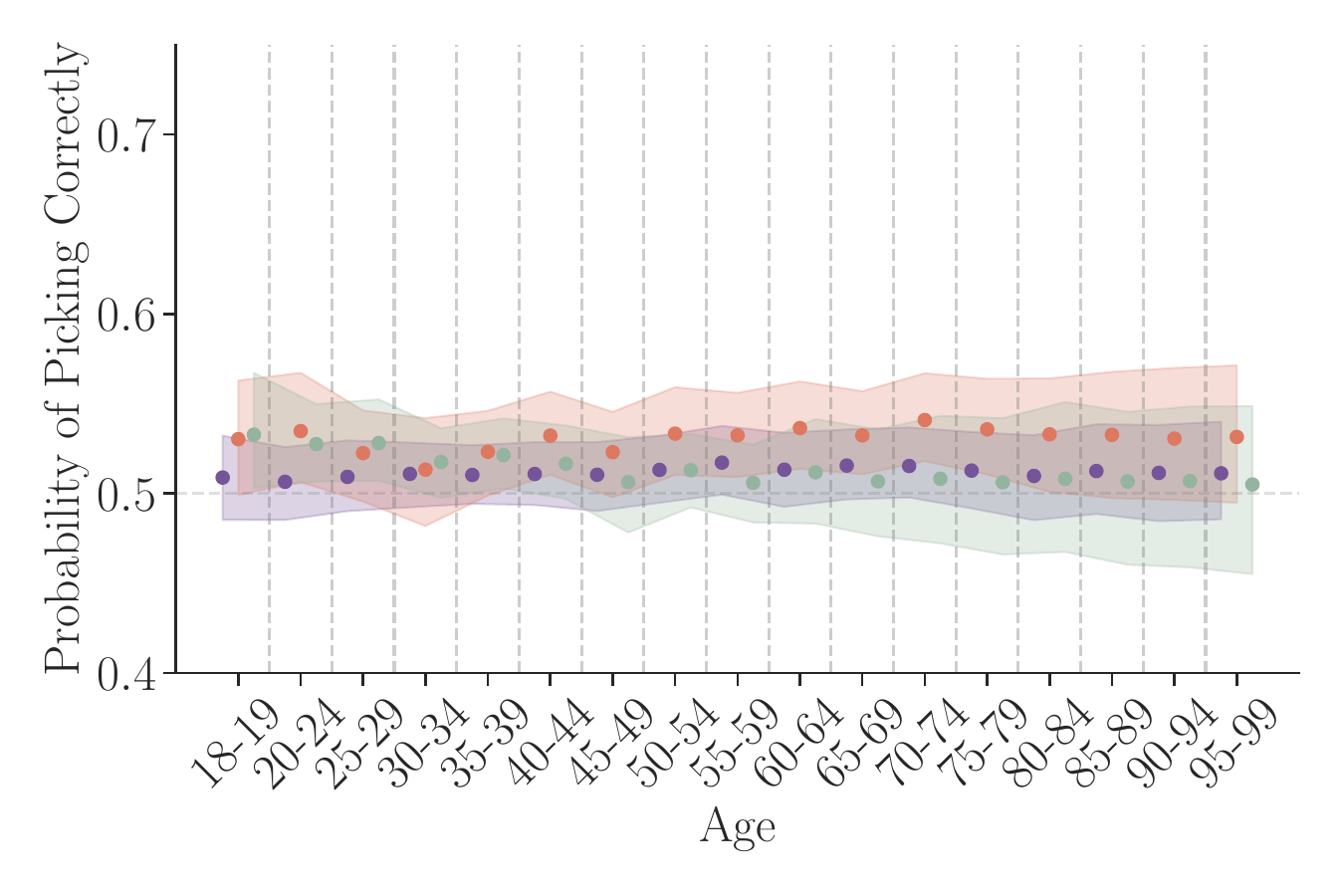}
         \caption{Text}
        \label{fig:demographic:age:text}
     \end{subfigure}
    \caption{\textbf{Predicted Posterior Probabilities split by Age Bracket} We plot the predicted posterior probability of picking the correct result by media type. 
    We plot the probabilities separated into the different age brackets.
    Note that we only show the probability range of $0.4$ to $0.75$ to better highlight the results.
    $0.5$ corresponds to random guessing, higher values indicate a higher accuracy in picking the correct result.
    }
    \label{fig:demographic:age}
\end{figure} \begin{figure}[t!bp]
     \centering
     \begin{subfigure}[b]{\linewidth}
         \centering
         \includegraphics[width=\linewidth]{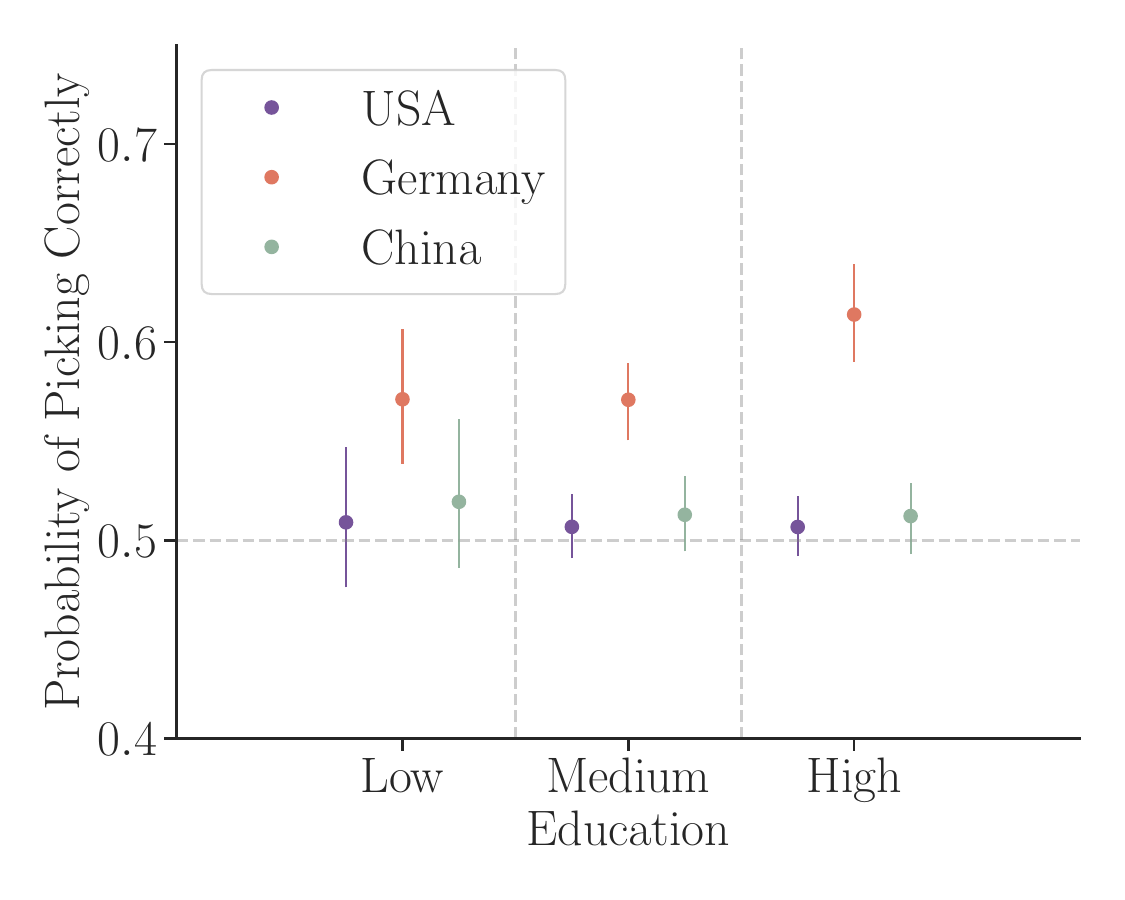}
         \caption{Audio}
        \label{fig:demographic:edu:audio}
     \end{subfigure}
     \vfill
     \begin{subfigure}[b]{\linewidth}
         \centering
         \includegraphics[width=\linewidth]{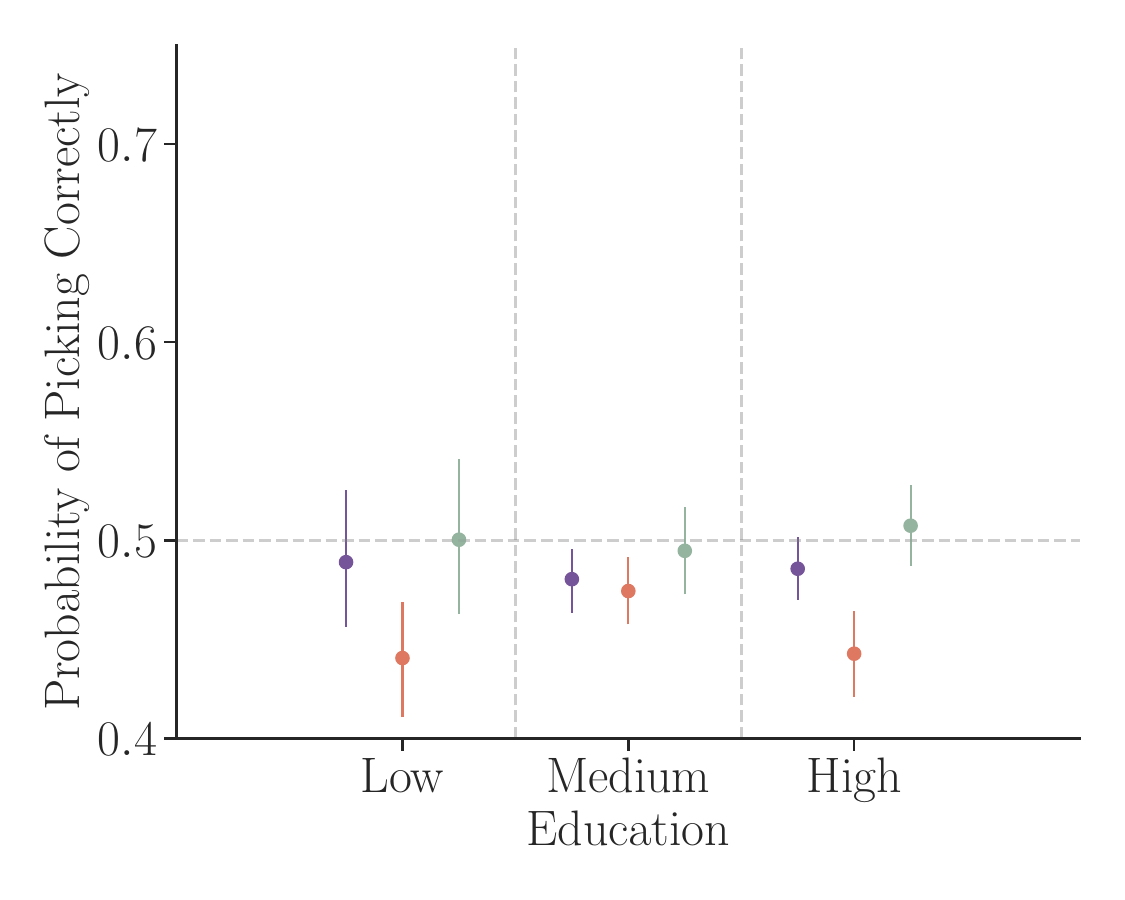}
         \caption{Image}
        \label{fig:demographic:edu:image}
     \end{subfigure}
    \caption{\textbf{Predicted Posterior Probabilities split by Education Levels} 
    We plot the predicted posterior probability of picking the correct results by media type. 
    We plot the results separated into the different education groups.
    Note that we only show the probability range of $0.4$ to $0.75$ to better highlight the results.
    $0.5$ corresponds to random guessing, higher values indicate a higher accuracy in picking the correct result.
    }
    \label{fig:demographic:edu}
\end{figure} 

When investigating the influence of demographic variables, we find that they only have marginal influence on Chinese and US participants.
The biggest influence can be observed for German participants, who were presented with audio data.
This indicates that when participants only guess, the influence of any one demographic variable tends to disappear.

We plot the influence of age split across media types in Figure~\ref{fig:demographic:age}.
For readability, we plot the overall predicted posterior probability of picking the correct result.
We compute this probability by keeping all other influences at average and then sampling the Gaussian Process we use to model the influence of age.
The biggest impact can be spotted for German participants presented with audio files, where we can observe a steady decrease towards guessing  with progressing age (mean predicted probability decreasing from $68.58\%$ to $50.99\%$).
Other less prominent trends are noticeable for German participants in the image condition (a decrease from $47.23\%$ to $44.15\%$) and for Chinese participants in the text condition ($53.28\%$ to $50.50\%$).

In Figure~\ref{fig:demographic:edu}, we plot the influence of the education parameter.
We observe a slight increase in the mean probability for German participants in the audio condition.
However, all of the posterior distributions still overlap.
We also observe that the distributions reflecting lower education levels exhibit a higher variability, reflecting the general trend of online panels to not accurately sample these participants (\cf~Section~\ref{sec:discussion}).

\begin{insightbox}
\finding{
Overall, we find that demographic variables matter---until they do not.
When machine-generated media is still lagging behind in terms of quality, demographic variables can influence the rate of detection.
In our case, we observe an influence of the age of participants on the detection accuracy for audio files.
However, once the quality improves, the influence of the demographic variables collapses, and the probability of picking the correct result becomes almost the same across the different age/education~groups.
}
\end{insightbox}

\subsection{RQ3: Which Cognitive Factors Influence the Identification Accuracy?}
\label{sec:analysis:acc:personal}

\begin{table*}[t]
    \centering
    \caption{\textbf{Predictor Posterior Means and 89\,\% Prediction Interval}
    We display the posterior means and 89\,\% PI for different predictors per media categories.
    We highlight predictors where more than $95\%$ of the posterior distribution falls either above or below zero in \textbf{bold}.
    }
    \resizebox{\linewidth}{!}{\begin{tabular}{@{}lrrrrrrrr@{}}
             & \multicolumn{2}{c}{Audio} && \multicolumn{2}{c}{Image} && \multicolumn{2}{c}{Text} \\
             \rule{0pt}{2.4ex}& \multicolumn{1}{c}{Human} & \multicolumn{1}{c}{Machine} & & \multicolumn{1}{c}{Human} & \multicolumn{1}{c}{Machine} & & \multicolumn{1}{c}{Human} & \multicolumn{1}{c}{Machine} \\
             \midrule
            AHS & -0.04 [-0.07, -0.00] & \textbf{0.05 [\phantom{-}0.02, \phantom{-}0.09]} & & \textbf{-0.05 [-0.09, -0.01]} & 0.00 [-0.04, \phantom{-}0.04] & & \textbf{0.04 [\phantom{-}0.00, \phantom{-}0.07]} & \textbf{-0.04 [-0.08, -0.01]} \\
            CRT & \textbf{-0.14 [-0.24, -0.04]} & \textbf{0.17 [\phantom{-}0.07, \phantom{-}0.27]} & & \textbf{-0.22 [-0.31, -0.11]} & \textbf{0.17 [\phantom{-}0.08, \phantom{-}0.28]} & & \textbf{-0.09 [-0.19, -0.01]} & \textbf{0.16 [\phantom{-}0.07, \phantom{-}0.25]} \\
            FAM & \textbf{0.85 [\phantom{-}0.48, \phantom{-}1.53]} & \textbf{-0.68 [-1.32, -0.36]} & & \textbf{-0.36 [-0.47, -0.27]} & \textbf{0.22 [\phantom{-}0.11, \phantom{-}0.32]} & & \textbf{-0.30 [-0.38, -0.22]} & \textbf{0.18 [\phantom{-}0.10, \phantom{-}0.26]} \\
            GTS & \textbf{0.16 [\phantom{-}0.12, \phantom{-}0.20]} & \textbf{-0.11 [-0.15, -0.07]} & & \textbf{0.11 [\phantom{-}0.07, \phantom{-}0.14]} & \textbf{-0.11 [-0.14, -0.07]} & & \textbf{0.09 [\phantom{-}0.05, \phantom{-}0.12]} & \textbf{-0.10 [-0.13, -0.07]} \\
            NMLS CC & 0.05 [-0.00, \phantom{-}0.09] & 0.02 [-0.02, \phantom{-}0.07] & & \textbf{-0.14 [-0.20, -0.09]} & 0.04 [-0.01, \phantom{-}0.10] & & \textbf{0.10 [\phantom{-}0.05, \phantom{-}0.14]} & \textbf{-0.11 [-0.16, -0.07]} \\
            NMLS CP & \textbf{0.09 [\phantom{-}0.03, \phantom{-}0.14]} & \textbf{-0.19 [-0.25, -0.14]} & & \textbf{0.08 [\phantom{-}0.01, \phantom{-}0.14]} & \textbf{-0.08 [-0.15, -0.01]} & & \textbf{0.13 [\phantom{-}0.07, \phantom{-}0.19]} & \textbf{-0.10 [-0.15, -0.04]} \\
            NMLS FC & \textbf{-0.07 [-0.11, -0.02]} & 0.03 [-0.01, \phantom{-}0.08] & & \textbf{0.22 [\phantom{-}0.17, \phantom{-}0.27]} & \textbf{-0.23 [-0.28, -0.17]} & & -0.03 [-0.08, \phantom{-}0.01] & 0.03 [-0.01, \phantom{-}0.08] \\
            NMLS FP & -0.02 [-0.08, \phantom{-}0.03] & 0.02 [-0.03, \phantom{-}0.08] & & -0.04 [-0.11, \phantom{-}0.02] & \textbf{0.07 [\phantom{-}0.00, \phantom{-}0.14]} & & -0.02 [-0.07, \phantom{-}0.04] & -0.02 [-0.08, \phantom{-}0.03] \\
            PO & 0.02 [-0.11, \phantom{-}0.14] & \textbf{-0.11 [-0.22, \phantom{-}0.00]} & & \textbf{-0.41 [-0.52, -0.30]} & \textbf{0.26 [\phantom{-}0.15, \phantom{-}0.38]} & & 0.08 [-0.02, \phantom{-}0.19] & \textbf{-0.15 [-0.26, -0.05]} \\
        \end{tabular}
    }
    \label{tab:acc:variables}
\end{table*} 

Finally, we study the influence of different variables on the probability of classifying a given media correctly.
We again perform a regression analysis with the models selected in Section~\ref{sec:analysis:acc:realfake} and individually add our predictor variables presented in Section~\ref{sec:related:variables}.
The results are presented in Table~\ref{tab:acc:variables}.
We present the posterior means and $89\%$~\pred{}.
When the majority of the posterior distribution either falls below or above zero, the predictor is believed to be meaningful.
Negative scores indicate that the predictor decreases the probability of picking correct for the given category.

Note that these are prediction intervals and not confidence intervals, as Bayesian inference does not infer point estimates but ranges of plausible values.
This aspect, in conjunction with hierarchical modelling, allows Bayesian methods to circumvent the problem of multiple comparisons found in Frequentist methods.
Thus, we do not correct for multiple comparisons, which also aligns with remarks in~\citet{gelman2012we}.

We decided to model the parameters individually for the following reasons:
Due to the link function, all terms in a generalized linear model will interact.
This is true whether the estimates are obtained by Frequentist or Bayesian methods.
Consequently, we cannot simply include all measured terms in a regression without running the risk of Simpson's Paradox (\cf~\citet{pearl2022comment}).
Intuitively, including terms in a regression can both reveal and hide causal relationships in your data.
Thus, whether to include a term in the regression analysis cannot be derived from the data but needs to be based on our understanding of how the data was created. Therefore, we choose to model each predictor separately.

We found three predictor variables that were robust across all media types:
First, participants with a higher general trust (\gts{}) have a higher chance of detecting human-generated content (audio: P($\beta > 0) = 1.00$; image: P($\beta > 0) = 1.00$; text: P($\beta > 0) = 0.99$).
On the flip side, it negatively affects their ability to classify machine-generated samples (audio: P($\beta < 0) = 1.00$; image: P($\beta < 0) = 1.00$; text: P($\beta < 0) = 1.00$).
Second, participants who achieved a higher \crt{} score are worse at recognizing human-generated media (audio: P($\beta < 0) = 0.98$; image: P($\beta < 0) = 1.00$; text: P($\beta < 0) = 0.95$) but are better at recognizing machine-generated media (audio: P($\beta > 0) = 1.00$; image: P($\beta > 0) = 0.99$; text: P($\beta > 0) = 0.99$).
Finally, \fami{} helps participants recognize human-generated audio data while hindering their ability to detect image and text data (audio: P($\beta > 0) = 1.00$; image: P($\beta < 0) = 1.00$; text: P($\beta < 0) = 1.00$).
The results are flipped for machine-generated data (audio: P($\beta < 0) = 1.00$; image: P($\beta > 0) = 1.00$; text: P($\beta > 0) = 0.99$).

Additionally, we found several predictors that showed up only for some media types:
We found support that holistic thinking (\ahs{}) helped participants discern human-generated text data (text: P($\beta > 0) = 0.96$).
The trend is reversed and stronger when classifying machine-generated data (audio: P($\beta > 0) = 0.98$; text: P($\beta < 0) = 0.97$).
We also detected a general trend that holistic thinking worsened the participants' performance on image data (both human- and machine-generated) (human: P($\beta < 0) = 0.92$; machine: P($\beta < 0) = 0.85$), but both distributions overlap with zero.
When participants leaned more towards materialism (\ie, more conservative values), it worsened their performance in detecting machine-generated audio (P($\beta < 0) = 0.95$) and text (P($\beta < 0) = 0.98$) condition, as well as correctly classifying human-generated images (P($\beta < 0) = 1.0$).
It did help participants to correctly classify machine-generated images (P($\beta > 0) = 1.0$).

Finally, when examining media literacy, we use the four subscales proposed in the original publication~\citep{koc2016development}:
First, functional consuming (\nmls{} FC), \ie,  being -- technically -- able to consume media, had a strong negative effect on detecting human-generated audio samples (P($\beta < 0) = 0.99$) and machine-generated image samples (P($\beta < 0) = 1.00$).
We also detect a strong positive effect on correctly classifying images made by humans (P($\beta > 0) = 1.00$).
Second, critical consumption (\nmls{} CC), \ie, the ability to conceive media messages as subjective instead of neutral, exhibits a positive effect for human-generated text data (P($\beta > 0) = 1.00$) and negative effects for both human-generated image data (P($\beta < 0) = 1.00$) and machine-generated text data (P($\beta < 0) = 1.00$).
Third, critical prosuming (\nmls{} CP), \ie, the ability to critically analyze self-created work, has a positive effect on human-generated audio and text data (audio: P($\beta > 0) = 0.99$; text: P($\beta > 0) = 1.00$) and the results are flipped for machine-generated data (audio: P($\beta < 0) = 1.00$; text: P($\beta < 0) = 1.00$).
Finally, functional prosuming, \ie, the ability to technically construct media, only showed a positive effect on detecting machine-generated images media (P($\beta > 0) = 0.95$).

\begin{insightbox}
\finding{
We found that~\gts{}, \crt{}, \fami{}, and \nmls{} CP  consistently showed up across all media types and \ahs{}, \nmls{} FC, and \nmls{} CC for one or more media types.
A key observation is that while they often have strong influences, they counteract each other, e.g., a higher \crt{} score predicts better performance for recognizing machine-generated media but at the same time hinders the ability to detect human-generated samples.
We hypothesize that these observations match those in Section~\ref{sec:analysis:acc:realfake:demographic}, i.e., that AI-generated media is already at a level where people are mainly guessing, thus the influence of these variables collapses.
This highlights the need for more research into automatically detecting machine-generated media.
}
\end{insightbox}
 \section{Discussion}
\label{sec:discussion}

With this work, we aim to gain a better understanding of the current state of people's ability to detect AI-generated media in different countries.
Overall, the results of our study indicate that machine-generated media has become almost indistinguishable from ``real'' media.
This is especially true in the image domain, where participants from all countries performed even worse than random guessing.

This trend is not fully universal.
We found indications that German audio data still lacks behind.
Intuitively, one might suspect that Western media would be of higher quality, since it is more often used for research purposes.
For example, the original training set for GPT-3~\citep{brown-20-language} (our text model) does not include much Chinese text ($0.16\%$) when compared to English ($93.69\%$) (number of documents).

While there was a statistical difference between how Chinese and US participants rated machine-generated media, we could not detect the same difference between accuracy ratings (\cf{} Section~\ref{sec:analysis:description}).
When we analyzed this phenomenon in more detail, we found a consistent pattern that Chinese and US participants seem to resort to random guessing (\cf{} Section~\ref{sec:analysis:description} and Section~\ref{sec:analysis:acc:realfake}).
However, they did perform differently when looking at human- and machine-generated media separately, which explains the difference in ratings.
This suggests that machine-generated English media is not that far ahead as typically believed in the machine learning community~\citep{rae2021scaling}.

We can further connect our results to prior research by studying key predictors based on our literature review.
The most prominent are \crt{}, \fami{}, and \gts{}, which all have been found to be significant across all media types:
These findings extend previous works, which have found \gts{} and \crt{} to be significant factors when predicting the likelihood of accepting friend requests from machine-generated LinkedIn profiles and the sharing of political deepfakes, respectively.
Our results indicate that these singular findings might generalize to multiple media types and should be closely studied in the future.
Partly in line with related work, we found conservatism (in our case materialistic values) to negatively influence the detection of artificially generated text and audio files~\citep{Calvillo_2020}.
Participants holding more materialistic values were worse at identifying forged images.

Contrary to our preregistration, these findings did not show up along the lines of Western (traditionally associated with more analytic thinking) and Asian (more holistic thinking) cultures.
Additionally, while \ahs{} showed up as a significant predictor in our regression analysis, the results were somewhat inconclusive.
This highlights the importance of more cross-cultural research in this area.

 \section{Limitations}
\label{sec:limitations}

In this study, we investigated the ability of humans to identify human- and machine-generated media types detached from a particular context.
We believe that our results serve as a basis for people's general abilities to detect fake content across different media types and countries.
However, context might help or hinder people in their detection abilities.
We can only speculate that participants might perform worse at detecting AI-generated media in a real-world scenario, as they were told in this study that some media were fake and were instructed to identify human- and machine-generated media.
Therefore, future work should take context into consideration and/or study detection in real-world scenarios.
It has to be noted that the conclusions drawn from our results must take the lack of context into account.

Generally speaking, it is difficult for online panels to reach people with low education and older people (see~\cite{redmiles-19-mturk-generalization,tang-22-generalize}).
Thus, we were not able to fully meet all representative quotas, especially for China. 
However, as generated media are more likely to be displayed on the Internet, we believe our sample still meets the target group, and we considered representative shortcomings in our analysis (see Section~\ref{sec:analysis:acc:realfake}).
Nevertheless, we believe it would be worthwhile to sample especially the aforementioned underrepresented groups in future work.  

Furthermore, we could not accurately account for control variables, as our panel provider could not provide us with accurate information on which devices our participants used.
Additionally, the majority of participants in the audio group did not indicate which device they used for listening to the audio files.  
Thus, we abstain from entering these variables into the regression.

Finally, there is an expiration date to our conclusions.
The field of generative modelling is one of the most active in the machine learning community.
For example, during our study, diffusion models have almost completely replaced GAN-based methods.
These models are often paired with large-language models, generating impressive art pieces from simple text prompts.
While their use case is confined to generative artworks at the moment, one can suspect they will be used for malicious use cases in the future.
In the same vein, there have been several projects in recent years that address the problem of underrepresented languages or speech corpora~\cite{rust2021tokenizer,limisiewicz2022balance,zhang2019learning}.
 \section{Conclusion}
\label{sec:conclusion}

In this paper we present the first cross-country, cross-media study on people's ability to detect AI-generated media.
We found that media generated with current state-of-the-art methods has become virtually indistinguishable from ``real'' media.
Across all countries and all media types, people rated AI-generated samples as more likely to be produced by a human than a machine.
Additionally, our statistical analysis 
showed that participants mostly guessed randomly, and it was very challenging for them to decide which stimuli were machine-generated and what clues to look for.
Further, we found that generalized trust, cognitive reflection, familiarity, holistic thinking, political orientation, and some sub-scales of News Media Literacy had an effect on the ability to discern human- and machine-generated media.

Our results clearly show that machine-generative media are indistinguishable from real media. 
Since perfect technical detection seems unattainable, we argue that future research should not focus on how to avoid generative AI but rather, how to live with it.
One relevant aspect is the inclusion of context: People might perceive machine-generated media differently when they see it in news articles, social media posts, or advertisements.
In addition, research should explore which applications of AI are acceptable to humans and which are not.
For example, are people okay with automatically generated summaries of product reviews? Generated news articles? Chatbots for therapy?
We believe that it is the responsibility of policymakers to proactively regulate the use of generated media based on scientific evidence to take ethical and societal aspects into account.
Only careful legislation that takes human perception and their values into account can mitigate the harmful effects of artificially generated media, without hindering its positive impacts.
Whatever the future might hold, generative AI is here to stay, and we hope that this study can serve as a wake-up call for increased human-centered research into these methods.

\section*{Acknowledgments}
This work was supported by the Deutsche Forschungsgemeinschaft (DFG, German Research Foundation) under Germany's Excellence Strategy -- EXC-2092  \textsc{CaSa} -- 390781972 and the German Federal Ministry of Education and Research under the grants UbiTrans (16KIS1900) and AIgenCY (16KIS2012).

\printbibliography
\clearpage

\begin{appendices}
    \section{Statistical Analysis}
\label{app:model}

We model the number of times $Y_i$  a participant $i$ correctly judged a media out of all his trials $N_i$, as a binomial distribution:
$$
Y_i \sim Binomial(N_i, p_i) \enspace,
$$
where we parameterize the probability $p_i$ by several different models.
In our notion we use superscripts as names and subscripts as indexes; for example, in the expression $\delta^{age}_{c,a}$  the parameter is named $\delta^{age}$ and indexed by country $c$ and age $a$.
Our full-model (Age/Education/Time by Country) models the probability $p_i$ as:

\begin{equation*}
    \begin{aligned}
        logit(p_i) &= \alpha^{country}_{c} + \delta^{edu}_{c,e} + \delta^{age}_{c,a} + \delta^{time}_{c} X^{time}_{i} \\
        \\
        \alpha_{c}^{country} &\sim N(\mu^{country}, \sigma^{country}) \\
        \delta_{c,e}^{edu} &\sim N(\mu^{edu}, \sigma^{edu}) \\
        \delta_{c}^{time} &\sim N(\mu^{time}, \sigma^{time}) \\
        \\
        \mu^{country}, \mu^{edu}, \mu^{time} &\sim N(0, 1.5) \\
        \sigma^{country}, \sigma^{edu}, \sigma^{time} &\sim \lambda(1) \\
        \\
        \begin{pmatrix}
        	\delta^{age}_{c,1} \\
        	\delta^{age}_{c,2} \\
        	\cdots \\
        	\delta^{age}_{c,A} \\
        \end{pmatrix} &\sim 
        N\begin{bmatrix}\begin{pmatrix}
        	0 \\
        	0 \\
        	\cdots \\
        	0 \\
        \end{pmatrix}, K \\ \end{bmatrix}\\
        \\
        K_{x,y} &= \eta_{c}^2 \exp(-\rho_{c}^2 D^2_{x,y}) + \pi{x,y} \, \sigma_{c}^2 \\
        \eta_{c}^2 &\sim \lambda(\eta^{age}) \\
        \sigma_{c}^2 &\sim \sigma^{age} \\
        \rho_{c}^2 &= \frac{1}{2} l_{c}^2 \\
        l_{c} &\sim HalfNormal(l^{age})  \\
        \\
        \eta^{age} &\sim \lambda(2) \\
        l^{age} &\sim HalfNormal(5)  \\
        \sigma^{age} &\sim \lambda(1)
    \end{aligned}
\end{equation*}

Intuitively, the model assumes that while each country has some freedom in the effect of the education/age levels, they should correlate across countries.
Thus, our model uses partial-pooling to model these correlations.
More specifically, the varying intercept $\alpha^{country}_c$ per country is used to model general differences between the countries (e.g., quality of different generative models).
We include two correction terms: the term $\delta^{edu}_{c,e}$ for each education level per country and the term $\delta^{time}_{c} X^{time}_{i}$ for the median time taken in the survey.
When modelling the age of our participants, we would expect that similar age brackets (e.g., people around the ages 25-30 and 30-35) are characterized by more similar behavior than age brackets that are far apart (e.g., 20-25 and 60-65).
We divide the age of our participants into 17 categories, spanning 5 years each, and used a Gaussian process to model the interactions between the different levels with an age-specific offset $\delta^{age}_{c,a}$.
The offset $\delta^{age}_{c,a}$ is drawn from a multivariate Normal distribution with mean zero and a covariance matrix defined by the kernel function $K$.
The covariance between any age pair $x,y$ equals the maximum covariance $\eta^2$, which is reduced at the rate $\rho^2$ by the squared distance between the two, $D_{x,y}^2$. There is an additional covariance parameter $\sigma^2$ with a dummy variable $\pi{x,y}$ which indicates that $x=y$. 
Thus, it expresses the additional covariance within each age group.
We include our predictor variables as fixed terms and model them either as continuous or ordinal variables~\citep{mcelreath2020statistical}.
During the analysis we standardize continuous scores and encode the Inglehart index, the CRT and knowledge score as ordinal variables.

All models were developed using pymc 4.3.0~\citep{Salvatier2016} with aesara 2.8.7~\citep{aesara} and jax 0.3.24~\citep{jax2018github} as the backend.
For each model, we started development on synthetic data, using prior predictive checks.
When using real data, we fit the parameters of all models with MCMC using the NUTS sampler for 4,000 iterations (2,000 warm-up).
Model fit was confirmed by inspecting $\hat r$-values.
All analyses were run on a server running Ubuntu 18.04.6 with an Intel Xeon Gold 6130 CPU and 128GB RAM.

     \section{Survey Design}
\label{app:survey}
\noindent Below we provide a screenshot illustrating our survey design (exemplary for the image condition).\smallskip

\begin{center}
      \includegraphics[width=\linewidth]{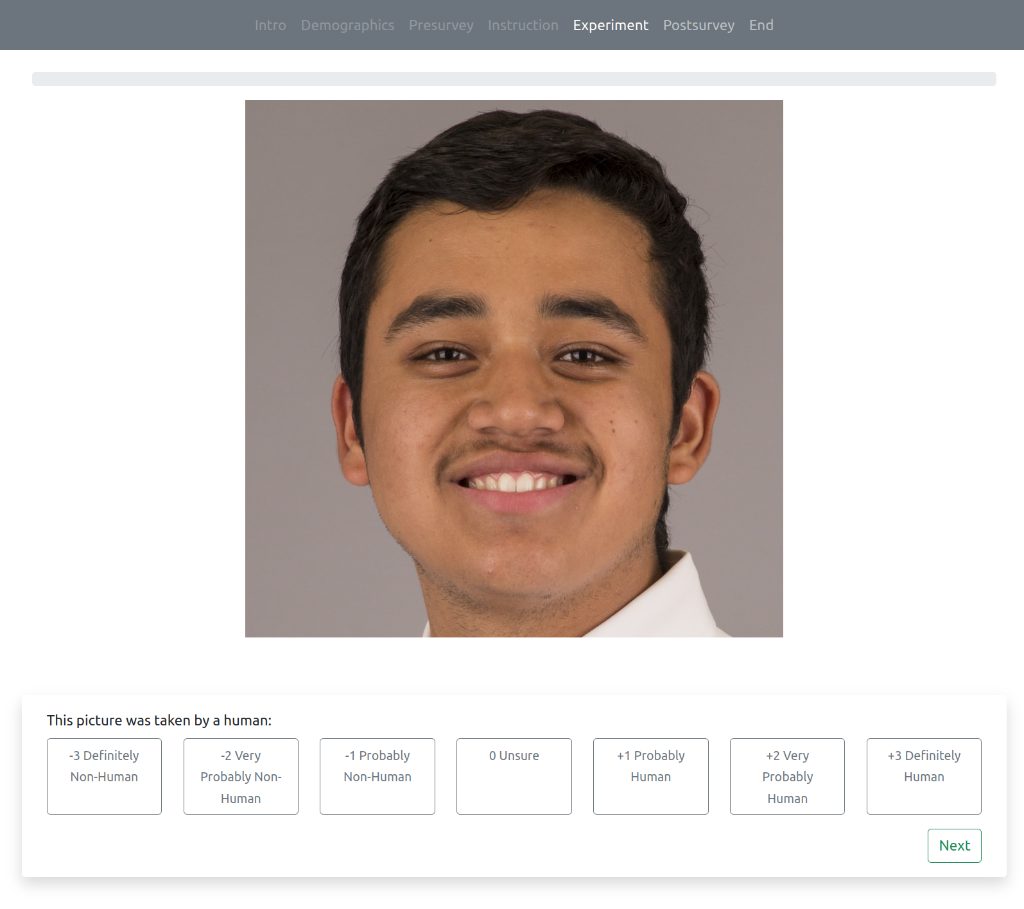}
\end{center}
     \onecolumn
    \section{Stimuli}
\label{app:stimuli}

\noindent Below we report exemplary stimuli for the three types of media used in the study. For speech, we show the audio files in the frequency domain. The full set of all stimuli is shared together with the preregistration 
(\href{https://osf.io/xy6v5}{osf.io/xy6v5}). 
\vspace{-0.5em}

\begin{center}
    \resizebox{\linewidth}{!}{\begin{tabular}{p{1.5cm}cc} 
    \toprule
            & Human-Generated 
            & Machine-Generated  \\
    \midrule
    & \phantom{a} & \\
    Image   & \parbox[c]{3.5cm}{\includegraphics[height=3.5cm]{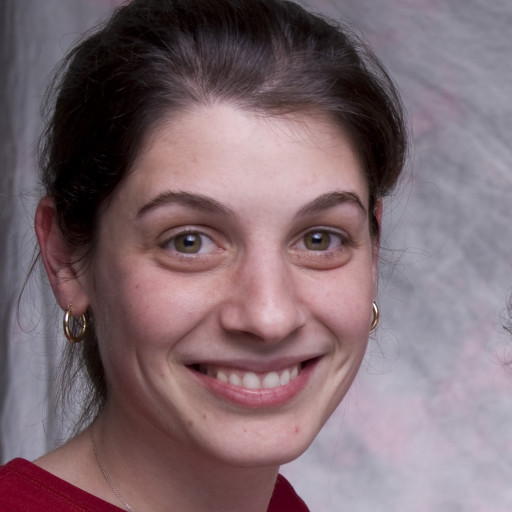}} 
            & \parbox[c]{3.5cm}{\includegraphics[height=3.5cm]{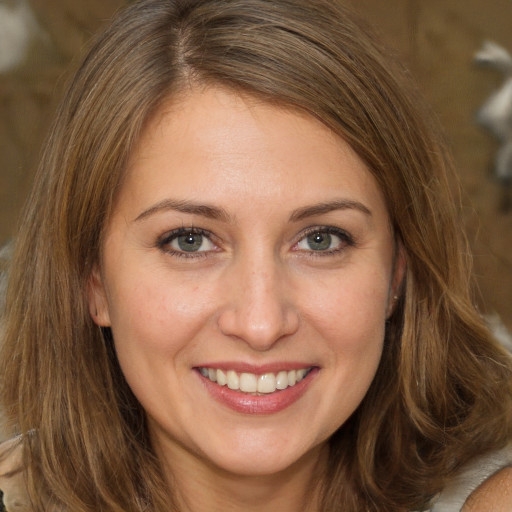}} \\
    & \phantom{a} & \\
Audio   & \parbox[c]{7cm}{\includegraphics[trim=6 3 6 3, clip, width=7cm]{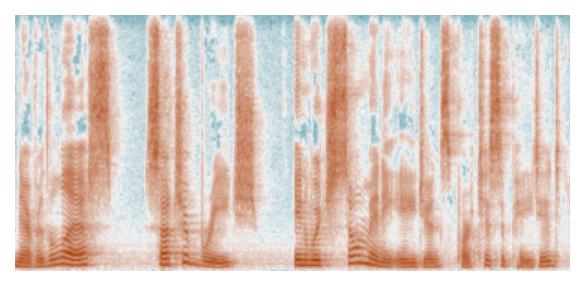}} 
            & \parbox[c]{7cm}{\includegraphics[trim=6 3 6 3, clip, width=7cm]{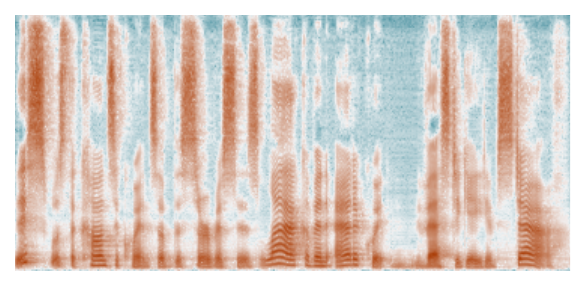}} \\
    & \phantom{a} & \\
    Text    & \parbox[c]{8cm}{Russia escalated its bombardment of the Ukrainian capital and launched new assaults on the port city of Mariupol, making bloody advances on the ground as Ukraine's president prepared Wednesday to make a direct appeal for more help in a rare speech by a foreign leader to the U.S. Congress. As the invasion entered its third week, Ukrainian President Volodymyr Zelenskyy suggested there was still some reason to be optimistic negotiations might yet yield an agreement with the Russian government.} 
            & \parbox[c]{8cm}{A magnitude 7.3 earthquake struck off the coast of Fukushima on Wednesday evening, triggering a tsunami advisory and cutting power to more than 2 million Tokyo homes, authorities said. The quake knocked out power to about 1.9 million households in the capital region shortly after 6 p.m., NHK national television reported. That figure was later revised to just over 2 million households as crews repaired damage caused by an earlier blackout, it said.} \\
    & \phantom{a} & \\
    \bottomrule
    \end{tabular}}
\end{center}

 \end{appendices}

\end{document}